%% file: main.tex
\documentclass[]{bytedance}
\usepackage[toc,page,header]{appendix}

\usepackage{minitoc}
\usepackage{amsfonts}
\usepackage{amssymb}
\usepackage{tabularx}
\usepackage{listings}
\usepackage{xcolor}
\usepackage{cancel}

\usepackage{caption} 
\usepackage{tabulary,multirow,xspace}
\usepackage{fixmath,mathtools,nicefrac,mmstyle}
\usepackage{subcaption}
\usepackage{caption}
\usepackage{wrapfig} 
\usepackage[misc]{ifsym} 
\usepackage{colortbl}
\usepackage[table]{xcolor}

\usepackage{nicematrix}
\usepackage{wrapfig}
\usepackage{multicol}
\usepackage[most]{tcolorbox}
\usepackage{pifont}

\usepackage{graphicx}
\usepackage[export]{adjustbox}
\usepackage{booktabs}
\usepackage{afterpage}

\def\eg{\emph{e.g.}\xspace}

\def\vs{\emph{vs.}\xspace}

\definecolor{codegreen}{rgb}{0,0.6,0}
\definecolor{codegray}{rgb}{0.5,0.5,0.5}
\definecolor{codepurple}{rgb}{0.58,0,0.82}
\definecolor{backcolour}{rgb}{0.95,0.95,0.92}
\definecolor{boxblue}{RGB}{57,89,163}
\definecolor{boxbluebg}{RGB}{230,237,250} 
\definecolor{myblue}{RGB}{210, 225, 255}

\lstdefinestyle{mystyle}{
    backgroundcolor=\color{backcolour},   
    commentstyle=\color{codegreen},
    keywordstyle=\color{magenta},
    numberstyle=\tiny\color{codegray},
    stringstyle=\color{codepurple},
    basicstyle=\ttfamily\footnotesize,
    breakatwhitespace=false,         
    breaklines=true,                 
    captionpos=b,                    
    keepspaces=true,                 
    numbers=none,                    
    numbersep=5pt,                  
    showspaces=false,                
    showstringspaces=false,
    showtabs=false,                  
    tabsize=2
}
\definecolor{mygray1}{gray}{.95}
\definecolor{mygray2}{gray}{.9}
\definecolor{mygray3}{gray}{.95}
\usepackage{pifont}

\newlength\savewidth
\newcolumntype{x}[1]{>{\centering\arraybackslash}p{#1pt}}

\newcommand{\app}{\raise.17ex\hbox{$\scriptstyle\sim$}}

\renewcommand{\emph}[1]{\textit{#1}}

\usepackage{xcolor}
\input{macros}

\usepackage{listings}

\definecolor{commentgreen}{rgb}{0.1, 0.4, 0.1}
\definecolor{keywordblue}{rgb}{0.1, 0.1, 0.7}
\definecolor{stringred}{rgb}{0.7, 0.1, 0.1}

\lstdefinestyle{mystyle}{
    commentstyle=\color{commentgreen},
    keywordstyle=\color{keywordblue},   
    stringstyle=\color{stringred},
    basicstyle=\ttfamily\scriptsize, 
    breaklines=true,
    keepspaces=true,
    showstringspaces=false,
    frame=none,                     
    language=Python, 
}

\title{
    GaMMA: Towards Joint Global-Temporal Music Understanding in Large Multimodal Models
}

\author[1,2]{Zuyao You}
\author[2]{Zhesong Yu}
\author[2]{Mingyu Liu}
\author[2\dagger]{Bilei Zhu}
\author[2]{Yuan Wan}
\author[1]{Zuxuan Wu}

\affiliation[1]{Fudan University}
\affiliation[2]{ByteDance}

\contribution[\dagger]{Project lead}

\abstract{
  In this paper, we propose GaMMA, a state-of-the-art (SoTA) large multimodal model (LMM) designed to achieve comprehensive musical content understanding. 
  GaMMA inherits the streamlined encoder-decoder design of LLaVA, enabling effective cross-modal learning between music and language. By incorporating audio encoders in a mixture-of-experts manner, GaMMA effectively unifies both time-series and non-time-series music understanding tasks within one set of parameters.
  Our approach combines carefully curated datasets at scale with a progressive training pipeline, effectively pushing the boundaries of music understanding via pretraining, supervised fine-tuning (SFT), and reinforcement learning (RL).
  To comprehensively assess both temporal and non-temporal capability of music LMMs, we introduce MusicBench, the largest music-oriented benchmark, comprising 3,739 human-curated multiple-choice questions covering diverse aspects of musical understanding.
  Extensive experiments demonstrate that GaMMA establishes new SoTA in the music domain, achieving \(79.1\%\) accuracy on MuchoMusic, \(79.3\%\) on MusicBench-Temporal, and \(81.3\%\) on MusicBench-Global, consistently outperforming previous methods.
}

\date{May 1, 2026}

\checkdata[Project Page]{\url{https://geshang777.github.io/GaMMA/}}

\begin{document}
\maketitle

\input{sec/1_intro}
\input{sec/2_related}
\input{sec/3_method}
\input{sec/4_exp}
\input{sec/5_con}

\bibliographystyle{plainnat}
\bibliography{main}
\clearpage
\beginappendix

\input{sec/X_appen}

\end{document}

%% file: macros.tex
\usepackage{graphicx}
\usepackage{amssymb}
\usepackage{pifont}
\usepackage{floatrow}
\usepackage{amsmath} 
\usepackage{float}
\usepackage{wrapfig}
\usepackage{multirow}
\usepackage{tcolorbox}
\tcbuselibrary{breakable, skins, raster}
\usepackage{listings}

%% file: sec/1_intro.tex
\section{Introduction}
Recent years have witnessed the rapid development of large language models (LLMs), one of the core trends is to develop a multi-modal assistant aligned with human intent that can effectively complete various real-world tasks in the wild~\cite{liu2023visual,askell2021general,driess2023palm}.
To this end, the research community has shown growing interest in transferring language-only LLMs into large multimodal models (LMMs). Recent advancements mainly focus on how to develop an open-world visual assistant~\cite{liu2023visual,meng2024deepstack,you2025pix2cap}, enabling models to perceive and understand the visual world more like humans. However, human interaction with the world extends beyond vision and language; sound, particularly music, serves as a universal medium for humans.

 While current multimodal models increasingly incorporate audio as inputs~\cite{team2024qwen2,chu2023qwen}, they act more like an LLM coupled with an automatic speech recognition system, as they are overly focusing on the language content in audio, and overlook the intrinsic properties of music like melody. Recently, a few music-oriented large models~\cite{liu2024music,liu2023m,tang2023salmonn} have been proposed; however, their performance on music understanding benchmarks~\cite{weck2024muchomusic} remains suboptimal. Besides, they cannot effectively model the temporal dynamics of music. However, music is inherently a temporal sequence, where meaning emerges through the evolution of notes, rhythm, and structure over time. Without modeling these temporal dynamics, models cannot truly understand or reason about music.

\begin{figure}[t]
  \centering
  \includegraphics[width=0.73\linewidth]{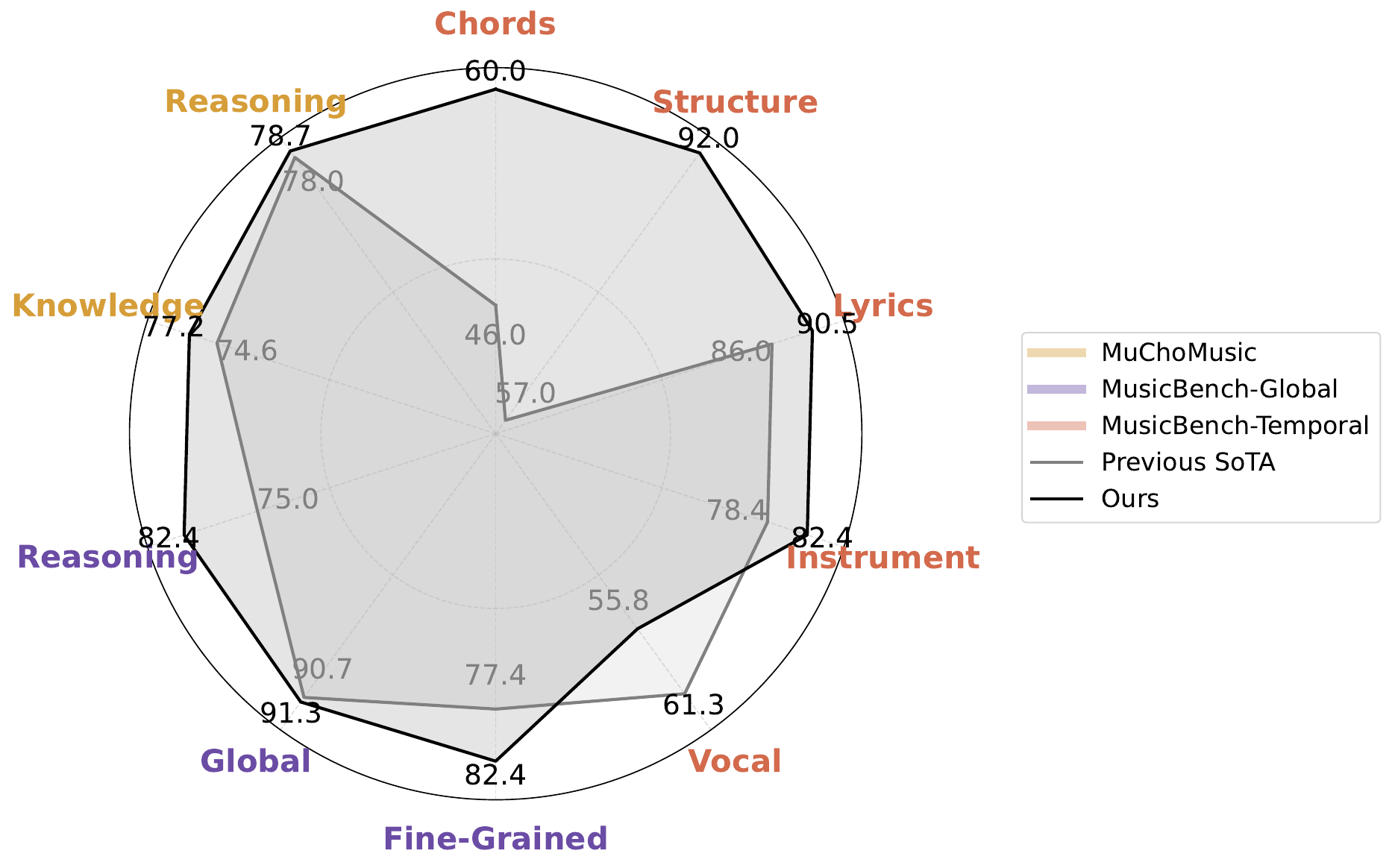}
  \caption{Comparison on several music understanding benchmarks. GaMMA-8B consistently achieves SoTA performance compared to the previous models with similar size. }
  \label{radar}
\end{figure}

To bridge this gap, we introduce GaMMA, \underline{G}eneral \underline{a}nd \underline{M}ultimodal \underline{M}usic \underline{A}ssistant, an all-round music LMM that can handle both time-series and non-time-series tasks within one set of parameters. Since music inherently contains both time-series elements (\eg, music structure) and semantic attributes (\eg, genre), GaMMA is explicitly designed to balance these complementary aspects. Unlike single-encoder architectures~\cite{chu2023qwen,liu2023m} that struggle to learn unified representations across these tasks, GaMMA employs dual specialized audio encoders, one for capturing time-series dynamics and another for broader musical content, both based on Whisper~\cite{radford2023robust} and augmented with MLP layers. These encoders operate as different experts, producing distinct embeddings which are then fused using a cross-attention-based extractor-injector mechanism~\cite{chen2022vision}. A gating module adaptively balances the contributions of each expert, while residual connections preserve original feature information. To support full-length audio inputs, music is segmented into \(30\)-second clips, tokenized proportionally, and processed with special boundary tokens.

We propose a three-stage training strategy to enhance the multimodal alignment, temporal understanding, and instruction-following capabilities of GaMMA. In the first stage, GaMMA is pretrained on large-scale paired data to establish both music-caption and music-lyric alignment. The second stage is supervised fine-tuning (SFT), integrating instruction-tuned and temporally annotated music data, injecting temporal awareness, and improving multi-turn dialogue capabilities. 
In the final stage, we introduce reinforcement learning (RL) to refine the performance of the model on complex music multi-choice tasks. In particular, we leverage Group Relative Policy Optimization (GRPO)~\cite{lu2024deepseek}, a recent proposed RL algorithm that has demonstrated strong potential in both visual~\cite{chen2025r1v,shen2025vlm,wang2025simplear} and audio~\cite{li2025reinforcement,xie2025audio} domains. We use both format and accuracy reward functions to guide the optimization process, enabling GaMMA to better navigate complex decision spaces over grouped reasoning trajectories.

Besides, existing music understanding benchmarks~\cite{weck2024muchomusic,sakshi2024mmaumassivemultitaskaudio} fail to explicitly assess the temporal capabilities of music LMMs. To comprehensively evaluate music LMMs, we introduce MusicBench, the largest and most comprehensive music-oriented benchmark, featuring 3,739 human-curated multiple-choice questions constructed via a rigorous annotation pipeline. 
We organize MusicBench into two subsets: MusicBench-Global and MusicBench-Temporal.
The MusicBench-Global contains \(2,741\) questions and is organized in a hierarchical manner to assess the model, ranging from global to fine-grained understanding, and from basic perceptual queries to complex reasoning tasks.
MusicBench-Temporal includes \(998\) questions and is specifically designed to evaluate the ability to reason about music over time. Each question requires the model to analyze an entire musical piece, interpret the properties of a specific time segment, and select the correct answer from multiple options.
MusicBench provides comprehensive coverage of both temporal and non-temporal aspects of music understanding, establishing a new standard for benchmarking music-language models.

We conduct extensive experiments to evaluate GaMMA against existing large audio multimodal models (LAMMs). As illustrated in Fig.\ref{radar}, GaMMA-8B achieves state-of-the-art (SoTA) performance on music understanding benchmarks like MusicBench and MuChoMusic~\cite{weck2024muchomusic}. Notably, in temporal understanding tasks, GaMMA significantly surpasses current LAMMs by a clear margin.

Our contributions can be summarized as follows:  
\begin{itemize}

    \item We introduce \textbf{GaMMA}, an all-around music LMM that bridges semantic and temporal understanding in music domain.

    \item We establish \textbf{MusicBench}, the most comprehensive evaluation benchmark for assessing both temporal and non-temporal ability of music LMMs. 
    
    \item Extensive experiments demonstrate that GaMMA consistently outperform existing LMMs, establish new SoTA in multiple benchmarks.

\end{itemize}

%% file: sec/2_related.tex
\section{Related Work}
\subsection{Large Audio Multimodal Models}
Recent advancements~\cite{liu2023visual,lu2024deepseek,meng2024deepstack} have expanded the capabilities of large language models (LLMs) to large multimodal models (LMMs), paving the way for Large Audio Multimodal Models (LAMMs). Early efforts~\cite{chu2023qwen,chu2024qwen2} primarily focused on aligning language with audio modalities such as speech and natural sounds; however, they paid limited attention to the musical domain.
Recently, several works~\cite{liu2023m,tang2023salmonn,alayrac2022flamingo} have emerged aiming to enhance the music understanding capabilities of LAMMs. While these models show promising progress, they often suffer from limited instruction-following ability and cannot process full-length music inputs. Besides, they are restricted to global-level tasks and lack explicit temporal modeling of music signals.
In this paper, we propose GaMMA, a SoTA LAMM that effectively supports both time-series and non-time-series music understanding tasks within a unified parameter space.

\subsection{Benchmarks for Large Audio Multimodal Models}
Evaluating the performance of LAMMs requires robust and domain-specific benchmarks. Early efforts such as CLAPScore~\cite{elizalde2023clap} primarily focused on text-audio similarity, offering limited insights into deeper semantic understanding. More recent benchmarks like AIR-Bench~\cite{yang2024air}, AISHELL-2~\cite{du2018aishell}, and CoVoST2~\cite{wang2021covost} aim to assess general audio-language alignment, but they pay little attention to the musical domain. Several recent benchmarks have been proposed specifically for music~\cite{weck2024muchomusic,liu2024music}, yet they only contain relatively simple temporal tasks, such as ``identifying the longest-playing instrument'', failing to explicitly evaluate the temporal reasoning capabilities of LAMMs in music understanding.
In contrast, we introduce MusicBench, a comprehensive benchmark designed for music-language models, featuring 3,739 human-crafted multiple-choice questions covering both temporal and non-temporal music tasks.

%% file: sec/3_method.tex
\section{Method}

\subsection{Modeling}

\begin{figure*}[t]
  \centering
  \includegraphics[width=\linewidth]{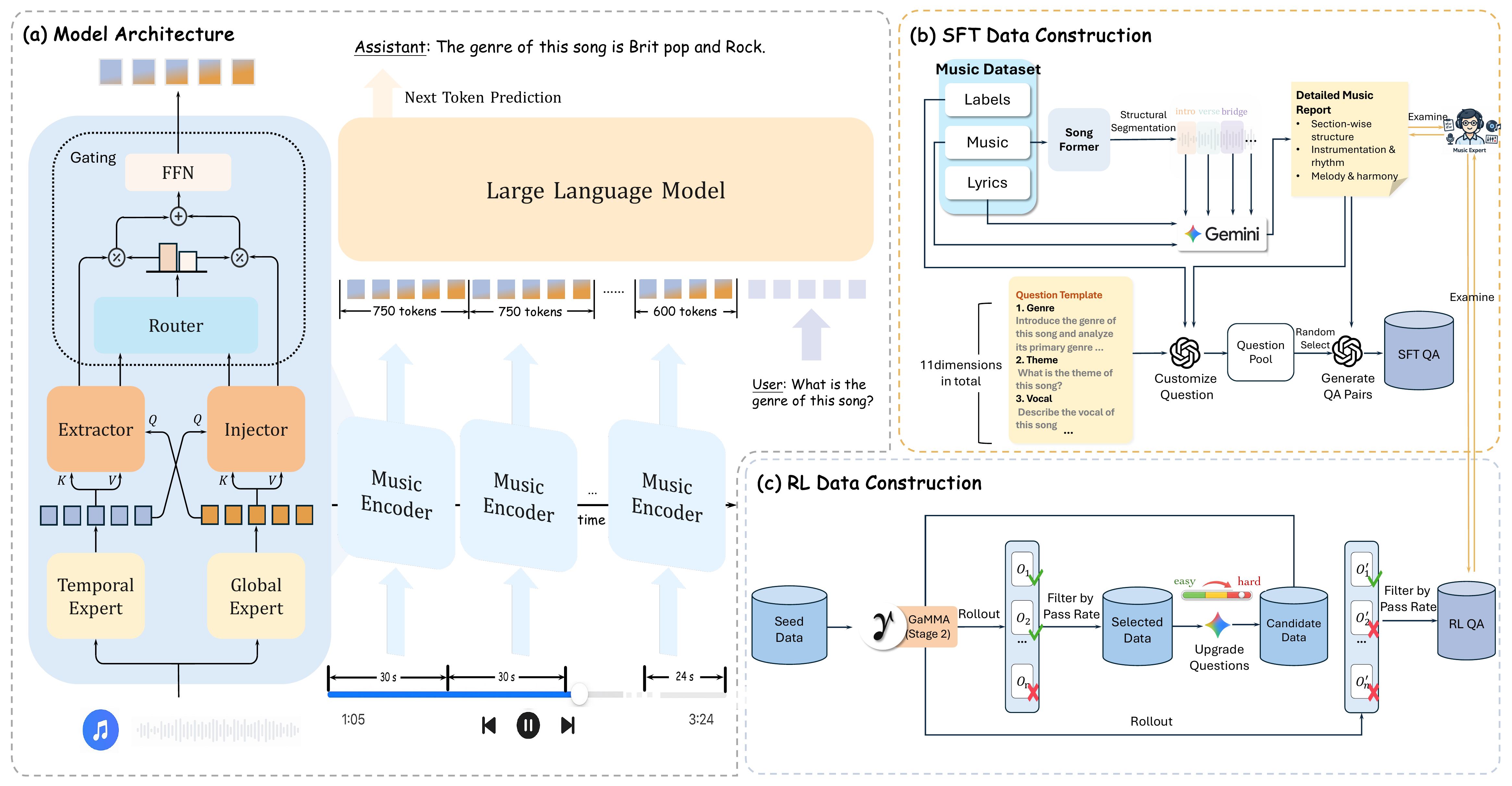}
  \caption{An overview of the model architecture and dataset construction pipeline of GaMMA.}
  \label{architechture}
\end{figure*}

The highlight of GaMMA lies in how to jointly handle the time-series tasks and non-time-series tasks, and how to support audio input of any length. As shown in Fig.~\ref{architechture}, GaMMA inherits the streamlined design of the encoder-decoder architecture from LLaVA~\cite{liu2023visual}. To begin with, we use the Whisper~\cite{won2024foundation} to embed audio and Qwen3~\cite{yang2025qwen3} as the LLM. The audio embedding from Whisper after multi-layer perceptrons (MLP) will be fed into the causal LLM for the next token prediction, where the attention mask is applied to force each text token to only see previous text tokens and the audio tokens. This prototype works well when processing the time-series data and non-time-series tasks separately, but the problem occurs when it comes to jointly training: when these two tasks are trained jointly, each task witnesses remarkable performance deterioration. As shown in Sec.~\ref{ablation_Sec}, we see this problem as the single encoder is not able to effectively handle these two tasks together, as these two tasks have different training objectives that acquire different embeddings from the encoder.

To address this challenge, we propose the Dual-encoder Fusion Network (DFN), which models heterogeneous encoders as complementary experts and fuses their representations in a unified framework. DFN consists of two Whisper-based encoders, each augmented with an additional MLP layer: a \emph{temporal expert}, fine-tuned on the temporal subset described in Sec.~\ref{traing_strategy}, and a \emph{global expert} for holistic musical content modeling. Each expert produces an embedding sequence \(e \in \mathbb{R}^{B \times S \times D}\), where \(B\), \(S\), and \(D\) denote the batch size, sequence length, and hidden dimension. Both encoders adopt the standard Whisper sinusoidal positional encoding; positional embeddings are instantiated independently and not shared between experts.

To fuse the expert embeddings, we adopt an extractor--injector mechanism inspired by ViT-Adapter~\cite{chen2022vision}, implemented via cross-attention without additional projection layers. Given embeddings \(e_1, e_2 \in \mathbb{R}^{B \times S \times D}\) from the temporal and global experts, the extractor and injector are defined symmetrically as:
\[
\hat{e}_1 = e_1 + \text{Softmax}\!\left(\frac{e_1 e_2^\top}{\sqrt{D}}\right) e_2, 
\]
\vspace{-7pt}
\[
\hat{e}_2 = e_2 + \text{Softmax}\!\left(\frac{e_2 e_1^\top}{\sqrt{D}}\right) e_1,
\]

Following cross-attention, we introduce a gating mechanism to adaptively route information from the two experts. Specifically, an MLP operates as a router on the feature-wise concatenation of the enhanced embeddings:
\[
\mathbf{g} = \sigma\bigl(\text{MLP}([\hat{e}_1, \hat{e}_2])\bigr),
\]
where \([\cdot,\cdot]\) denotes concatenation along the feature dimension. The gate \(\mathbf{g} \in \mathbb{R}^{B \times S \times D}\) is parameterized at the token level, enabling time-varying expert selection. To mitigate expert collapse, dropout is applied within the gating MLP during training. The fused embedding is computed as:
\[
e_{\text{fused}} = \mathbf{g} \odot \hat{e}_1 + (1 - \mathbf{g}) \odot \hat{e}_2.
\]
Finally, residual connections are preserved by combining the fused representation with the original expert outputs, followed by a feed-forward network:
\[
e_{\text{out}} = \text{FFN}([e_{\text{fused}},\, e_1 + e_2]).
\]

To process full-length audio, we segment the music into \(30\)-second chunks, each producing \(750\) tokens. For the final segment shorter than \(30\) seconds, no padding is applied; instead, the number of tokens is scaled proportionally to reflect its duration, providing the LLM with coarse length information. Tokens from all segments are concatenated to form the audio sequence, with special boundary tokens inserted to distinguish audio from text. The resulting audio tokens are jointly processed with text tokens by the LLM for next-token prediction.

\subsection{Dataset Curation}
\label{sec:dataset_curation}
\noindent\textbf{\textit{SFT Data Construction.}} During the SFT data construction stage, we first apply SongFormer~\cite{hao2025songformer} to segment each music track into fine-grained, second-level musical structures. As shown in Fig.~\ref{architechture}, the time-aligned music segments and corresponding lyrics are jointly fed into Gemini-2.5 Pro~\cite{team2024gemini} to generate detailed music analysis reports. The generated outputs then undergo rigorous verification by human experts, during which both temporal alignments and analytical content are filtered and rewritten to ensure semantic accuracy and reliability. 

To enable multi-turn music-oriented dialogue capabilities, we further construct conversation data based on the above music reports. We predefine a comprehensive set of question templates covering 11 musical dimensions (detailed in the Tab.~\ref{tab:rubric_1}-\ref{tab:rubrics_2}). To improve instruction-following ability and avoid overfitting to fixed templates, we use 
GPT-5.1 to rewrite and customize questions based on the predefined templates. High-quality question-answer pairs are then generated and randomly sampled to form the initial SFT training dataset.

\noindent\textbf{\textit{RL Data Construction.}}
To ensure that GaMMA receives sufficiently informative reward signals during the RL policy optimization stage, we adopt a data construction strategy inspired by SynthRL~\cite{wu2025synthrl}. We first generate an initial set of seed data using the same pipeline as in the SFT stage. Then we use GaMMA to estimate the pass rate of each question via Monte Carlo rollouts. Let the pass rate be denoted as \(Pass\). Questions with moderate difficulty, satisfying \(25\% \leq Pass < 1\), are selected as seed samples, avoiding questions that are either too easy or too difficult to provide effective training signals. We then perform question synthesis using Gemini-2.5 Pro. Given only the original question and the associated music clip, without access to the original answer, the model is prompted to generate candidate question variants that require deeper reasoning while preserving answer consistency. Finally, we apply GaMMA again for rollout-based verification, retaining questions that satisfy \( Pass - 25\% \geq Pass' \geq 25\% \), ensuring both answer validity and increased difficulty.

To mitigate data contamination, we perform audio-level overlap detection between all training sets and evaluation benchmarks. For each clip, we extract fixed-length embeddings using a frozen Whisper-based encoder and compute cosine similarity to all benchmark tracks. Any training sample with similarity greater than \(\tau = 0.95\) is discarded.

\subsection{Training Strategy}
\label{traing_strategy}

\noindent\textbf{\textit{Stage 1}: Multi-Modal Alignment Pretraining.} In this stage, we are working to establish audio-caption and audio-lyrics alignment in GaMMA. To this end, we utilize a comprehensive dataset of music-text pairs, combining large-scale open-source data, including Audioset-Cap~\cite{gemmeke2017audio} and MSD-Capsum~\cite{doh2023lp}.
The audio-caption data trains the model to connect musical attributes, while music-lyric pairs establish its automatic speech recognition (ASR) and lyric alignment capabilities. In this stage, we only keep the MLP trainable.

\noindent\textbf{\textit{Stage 2}: Supervised Fine-Tuning.}
In the second stage, we incorporate temporally annotated music reports and instruction-tuned question–answer pairs as described in Sec.~\ref{sec:dataset_curation}, to enhance the temporal understanding and multi-turn conversational abilities of GaMMA. At this stage, we also reintroduce a subset of the music captioning data from Stage 1. In addition, we leverage MIDI-synthesized audio from the Los Angeles~\cite{lev2024losangelesmididataset}, which provides precise annotations for chords, BPM, and key, to explicitly inject chord recognition capability into the model. To further strengthen general reasoning and dialogue skills of GaMMA, we integrate AudioSet-Strong~\cite{gemmeke2017audio} and several text-only datasets~\cite{amini2019mathqa,conover2023free,peng2023instruction,mitra2024orca,cobbe2021gsm8k,bai2024coig}. During this stage, GaMMA is fine-tuned with full-parameter trainable.

\noindent\textbf{\textit{Stage 3}: Reinforcement Learning.} To further improve the audio reasoning ability of GaMMA, we adopt Group Relative Policy Optimization (GRPO) in the final training stage, leveraging several thousand high-quality, reasoning-required multiple-choice questions construct using the pipeline in Sec.~\ref{sec:dataset_curation}. Given an input music and a query \( q \), GRPO samples a group of outputs sequences \( o_1, o_2, \ldots, o_G \) from the current policy \( \pi_{\theta_{\text{old}}} \), and updates the policy \( \pi_{\theta} \) by maximizing the following objective:

\begin{equation*}
\mathcal{J}_{GRPO}(\theta) =
\mathbb{E}_{o_{i} \sim \pi_{\theta_{old}}} \left[
\frac{1}{G}\sum_{i=1}^{G} \min \left(
\frac{\pi_{\theta}(o_{i} | t)}{\pi_{\theta_{old}}(o_{i} | t)} \hat{A}_{i},
\mathrm{clip} \left(
\frac{\pi_{\theta}(o_{i} | t)}{\pi_{\theta_{old}}(o_{i} | t)},
1 - \epsilon, 1 + \epsilon
\right)\hat{A}_{i}
\right)
- \beta \mathrm{D}_{KL} (\pi_{\theta} \| \pi_{ref})
\right]
\end{equation*}

The estimated advantage \( \hat{A}_{i} \) quantifies the relative utility of each sampled output within its group. Policy updates are clipped at threshold \( \epsilon \) to maintain training stability. Additionally, a KL divergence penalty between \( \pi_\theta \) and a reference policy \( \pi_{\text{ref}} \), weighted by a hyperparameter \( \beta \), is applied to regulate policy shifts. 
The reward function integrates accuracy and format fidelity: a correct answer receives an accuracy reward of \(1.0\), otherwise \(0\); for format, the reward is \(1.0\) only if the response is wrapped within \texttt{<answer>}...\texttt{</answer>} tags. Details on the training strategy are included in Sec.~\ref{sec:supp_strategy}.

\section{MusicBench}

\subsection{Pipeline and Quality Control}
\label{benchmark construction}
\begin{wrapfigure}{r}{0.55\columnwidth}
    \centering
  \includegraphics[width=0.9\linewidth]{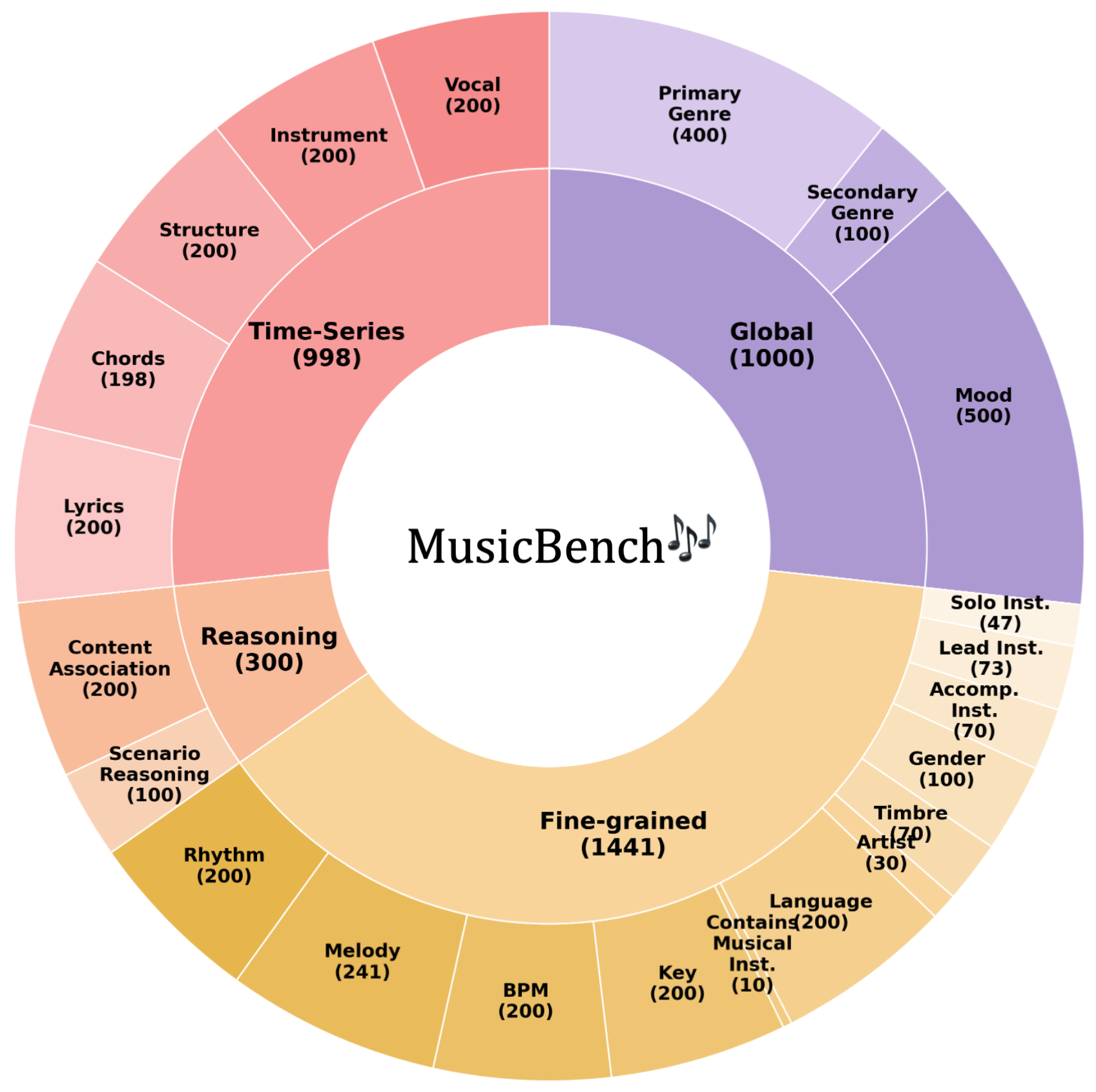}
  \caption{Hierarchical evaluation dimensions in MusicBench, comprising \(3,739\) manually labeled multiple choice questions.}
  \vspace{-10pt}
  \label{musicbench}
\end{wrapfigure}
\textbf{For the non-temporal subset}, we construct a comprehensive label pool covering a wide spectrum of musical attributes, including genre, mood, instrumentation, structure, and more. Annotators are instructed to use this label set to search for representative songs on YouTube and collect corresponding URLs. Each label is assigned a specific quota, ensuring that the dataset achieves a balanced distribution and avoids long-tail issues by maintaining adequate representation for each tag.

Once candidate audio tracks are identified, a team of highly qualified experts from renowned music academies will design question-answer pairs for different dimensions within MusicBench, as shown in Fig.~\ref{musicbench}. These experts create question prompts and answer options, ensuring that each incorrect option, apart from the correct one, is distinguishable from the correct answer and corresponds to other relevant labels in the pool. We standardize the format for both the questions and the answer options. For example, when referencing altered pitch keys, we consistently position accidentals to the right; thus, B-flat major would be written as ``\(B\)-\(flat\ major\)'', ``\(B^b major\)'', or ``\(B^b\):\(maj\)''. Similarly, altered pitches are denoted as ``\(C^\#5\)''. 

\textbf{For the temporal subset}, we aim for comprehensive coverage of all music-related tasks that require temporal understanding. Specifically, we evaluate models across five key dimensions: vocals, instruments, structure, chords, and lyrics. In these tasks, the model is required to analyze specific time segments of the music and identify the correct option from multiple choices based on temporal attributes. These tasks impose significant demands on the temporal modeling capabilities of music LMMs.

\subsection{MusicBench Statistics}
As illustrated in Fig.~\ref{musicbench}, MusicBench is a comprehensive evaluation framework designed for a detailed and nuanced analysis of musical understanding. 
It comprises a total of \(3,739\) manually curated multiple-choice questions. 
To facilitate targeted evaluation, MusicBench is organized into two primary subsets:
\begin{itemize}
    \item \textbf{MusicBench-Global}: A subset of \(2,741\) questions focused on non-temporal attributes such as genre, instrumentation, and key.
    \item \textbf{MusicBench-Temporal}: A subset of \(998\) questions specifically designed to assess reasoning over time, such as structure and chord progressions.
\end{itemize}
This hierarchical structure allows for a thorough assessment of a wide range of musical features. 
As detailed in Sec.~\ref{benchmark construction}, every question in MusicBench undergoes a rigorous dual-review process by two independent experts to ensure its accuracy and reliability. More details of MusicBench can be found in Sec.~\ref{sec:supp_musicbench}.

\begin{figure}[t]
    \centering
  \includegraphics[width=0.8\linewidth]{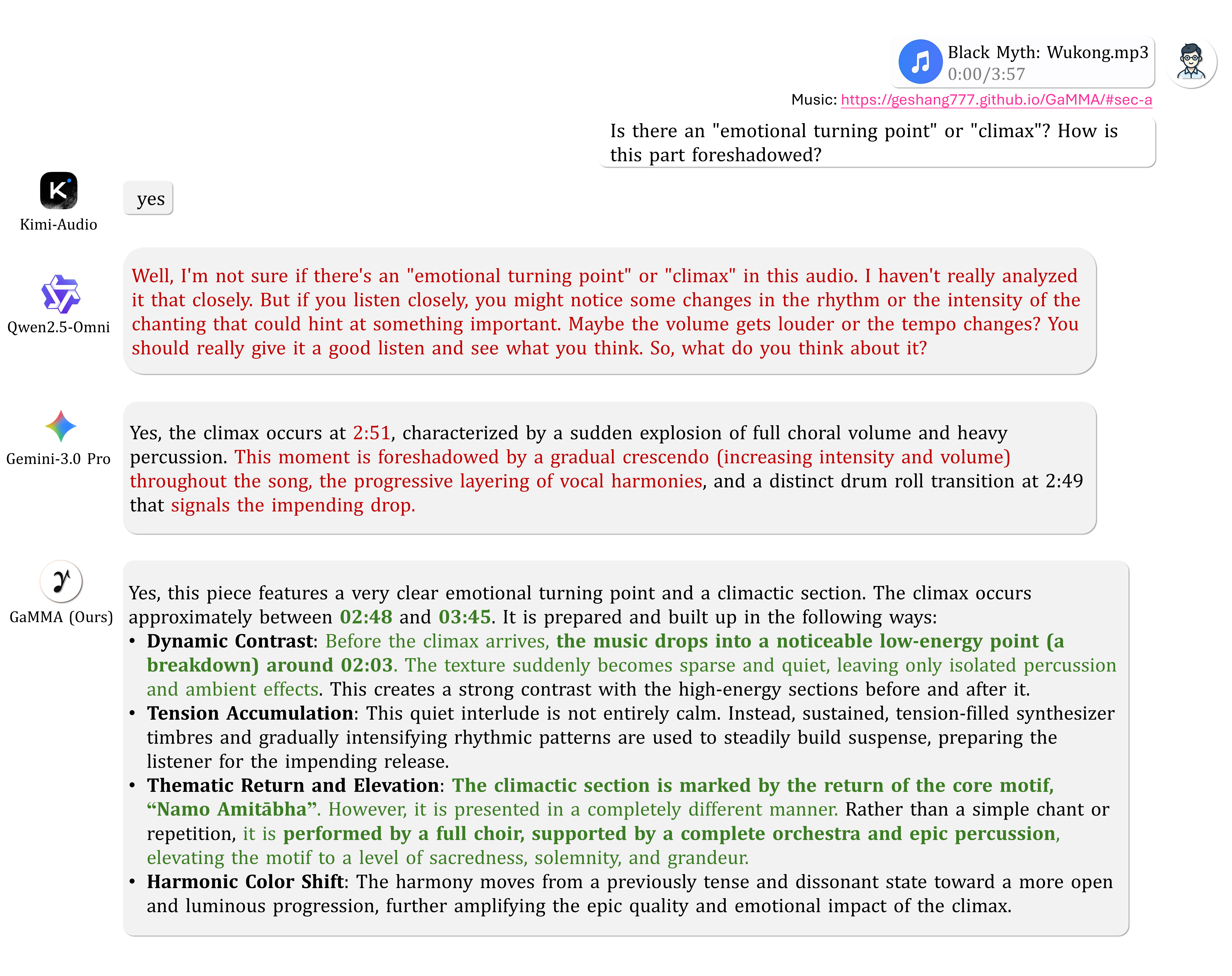}
  \caption{Our proposed GaMMA demonstrates superior semantic and temporal music understanding ability compared to Kimi-Audio~\cite{ding2025kimi} and Gemini-3.0 Pro. }
  \label{fig:com_vis}
\end{figure}

%% file: sec/4_exp.tex
\section{Experiments}
\subsection{Implementation Details} 

GaMMA is trained using a three-stage strategy as outlined in Sec.~\ref{traing_strategy}. We present two versions of our model: GaMMA-8B and GaMMA-14B, built upon Qwen3-8B and Qwen3-14B, respectively. 
In the pretraining stage, we use a batch size of \(256\) and a learning rate of \(1\times10^{-3}\). For SFT, we reduce the batch size to \(64\) and the learning rate to \(1\times10^{-5}\). Each stage is trained for one epoch with a warm-up period of \(0.03\) and a cosine learning rate scheduler. During the RL stage, the learning rate is set to \(10^{-6}\) and a batch size of \(4\). \(G\), \(\epsilon\), and \(\beta\) are set to \(4\), \(0.04\), and \(0.2\), respectively. Throughout training, we adopt DeepSpeed ZeRO-3~\cite{rasley2020deepspeed} and FlashAttention V2~\cite{dao2023flashattention} with mixed-precision BF16 training~\cite{kalamkar2019study}. Liger Kernel~\cite{hsu2025ligerkernel} is also applied to improve efficiency and memory usage. All experiments are conducted on \(16\) NVIDIA H100 GPUs with \(80\)G memory.

\subsection{Main Results}

\noindent\textbf{Experiments on MuChoMusic.} We choose the recently proposed benchmark MuChoMusic~\cite{weck2024muchomusic}, to compare our proposed GaMMA with current music LMMs. MuChoMusic is designed to assess models across two main evaluation dimensions: knowledge and reasoning. As shown in Table~\ref{tab:muchomusic_final_corrected}, GaMMA-8B achieves SoTA performance with an overall accuracy of \(78.0\%\), substantially outperforming all baseline models with similar parameter size. In reasoning-related tasks, GaMMA-8B achieves top scores across metrics like functional context (\(83.1\%\)) and temporal relations (\(90.0\%\)). Besides, it consistently outperforms strong baselines such as Qwen2.5-Omni~\cite{xu2025qwen2} in the knowledge tasks, with strong results in structure (\(78.9\%\) \vs \(68.0\%\)), melody (\(82.6\%\) \vs \(72.0\%\)), and instrumentation (\(78.1\%\) \vs \(78.0\%\)). To further explore the effect of model scaling, we upgrade the foundation model to Qwen3-14B. The resulting GaMMA-14B achieved an overall accuracy of (\(79.0\%\)), reaching a performance level comparable to Gemini-2.5 Pro, and significantly outperforming it on music-specific metrics such as melody understanding (\(82.6\%\) \vs \(70.7\%\)).

\begin{table*}[t]
  \caption{
    Experiment results on the MuChoMusic benchmark~\cite{weck2024muchomusic}, which assesses models across reasoning and knowledge dimensions. Reasoning includes: cultural context (CC), lyrics, temporal relations (TR), functional context (FC), genre, and mood. Knowledge encompasses: structure, harmony, melody, sound texture (ST), meter and rhythm (M\&R), performance (Perf.), and instrumentation (Inst.). The \textbf{best} result is in bold, and the \underline{second-best} is underlined.
  }
  \label{tab:muchomusic_final_corrected}
  \centering
  \setlength{\tabcolsep}{1.2mm} 

  \scalebox{0.8}{\begin{NiceTabular}{l*{15}{c}}
    \CodeBefore
      \columncolor{teal!7}{16}
    \Body
    
    \toprule

    {} & \multirow{2}{*}{\textbf{Param.}} & \multicolumn{6}{c}{\textbf{Reasoning}} & \multicolumn{7}{c}{\textbf{Knowledge}} & \multirow{2}{*}{\textbf{Acc.}} \\
    \cmidrule(rl){3-8} \cmidrule(rl){9-15}
    {} &{} & CC & Lyrics & TR & FC & Genre & Mood & Structure & Harmony & Melody & ST & M\&R & Perf. & Inst. & \\
    \midrule
        \textcolor{gray}{Gemini-2.0-Pro}  &\textcolor{gray}{-}        & \textcolor{gray}{63.7} & \textcolor{gray}{60.0} & \textcolor{gray}{65.0} & \textcolor{gray}{66.2} & \textcolor{gray}{59.1} & \textcolor{gray}{65.9} & \textcolor{gray}{68.4} & \textcolor{gray}{82.1} & \textcolor{gray}{73.9} & \textcolor{gray}{71.1} & \textcolor{gray}{57.8} & \textcolor{gray}{81.1} & \textcolor{gray}{64.6} & \textcolor{gray}{63.8} \\
        \textcolor{gray}{Gemini-2.5 Pro}  &\textcolor{gray}{-}             & \textcolor{gray}{83.3} & \textcolor{gray}{94.1} & \textcolor{gray}{94.6} & \textcolor{gray}{87.7} & \textcolor{gray}{78.9} & \textcolor{gray}{80.4} & \textcolor{gray}{100.0} & \textcolor{gray}{88.0} & \textcolor{gray}{70.7} & \textcolor{gray}{87.8} & \textcolor{gray}{83.3} & \textcolor{gray}{76.9} & \textcolor{gray}{78.0} & \textcolor{gray}{79.4} \\
        
        \textcolor{gray}{Gemini-3.0 Pro}  &\textcolor{gray}{-}             & \textcolor{gray}{85.7} & \textcolor{gray}{91.4} & \textcolor{gray}{90.0} & \textcolor{gray}{88.3} & \textcolor{gray}{76.8} & \textcolor{gray}{81.0} & \textcolor{gray}{84.2} & \textcolor{gray}{85.7} & \textcolor{gray}{84.8} & \textcolor{gray}{76.3} & \textcolor{gray}{82.2} & \textcolor{gray}{83.8} & \textcolor{gray}{76.2} & \textcolor{gray}{79.8} \\
    Mu-LLaMA~\cite{liu2024music} &7B & 25.0 & 40.0 & 22.5 & 23.4 & 33.3 & 25.4 & 15.8 & 21.4 & 37.0 & 31.6 & 28.9 & 29.5 & 33.3 & 30.6 \\
    M2UGen~\cite{liu2023m} &7B & 50.0 & 25.7 & 40.0 & 33.8 & 41.9 & 42.4 & 36.8 & 53.6 & 54.4 & 36.8 & 44.4 & 44.5 & 49.2 & 43.9 \\
    SALMONN~\cite{tang2023salmonn} &13B & 39.3 & 17.1 & 37.5 & 37.7 & 37.9 & 42.9 & 42.1 & 50.0 & 34.8 & 32.9 & 50.0 & 35.3 & 38.2 & 37.0 \\
    QwenAudio~\cite{chu2023qwen} &7B & 46.8 & 42.9 & 40.0 & 46.8 & 49.5 & 52.7 & 26.3 & 60.7 & 52.2 & 34.2 & 48.9 & 52.6 & 50.4 & 49.1 \\
    Qwen2Audio~\cite{chu2024qwen2} &7B & 42.9 & 54.3 & 65.0 & 61.0 & 53.3 & 64.4 & 52.6 & 64.3 & 50.0 & 64.5 & 71.1 & 66.5 & 64.6 & 61.5 \\
Kimi-Audio~\cite{ding2025kimi} & 8B
& \textbf{82.1} & 77.1 & 70.0 & 77.9 & 74.8 & 68.3 & 57.9 & 78.6 & 63.0 & 65.8 & 68.9 & 67.1 & 66.2 & 68.2 \\

Audio-Flamingo3~\cite{goel2025audio} & 8B
& \underline{78.6} & \textbf{91.4} & 85.0 & 68.8 & 66.7 & 75.6 & 57.9 & \textbf{92.9} & \underline{76.1} & \textbf{77.6} & \underline{76.7} & 73.4 & 73.9 & 73.4 \\

Qwen2.5-Omni~\cite{xu2025qwen2} & 8B
& 61.0 & 86.0 & \underline{88.0} & 82.0 & \underline{78.0} & \underline{78.0} & 68.0 & \underline{86.0} & 72.0 & 75.0 & 76.0 & 69.0 & \underline{78.0} & 76.0 \\

\midrule

\textbf{GaMMA-8B (ours)} & 8B
& 75.0 & \underline{88.6} & \textbf{90.0} & \underline{83.1} & 77.3 & 76.6 & \underline{78.9} & 78.6 & \textbf{82.6} & \underline{76.3} & \textbf{77.8} & \underline{78.0} & \textbf{78.1} & \underline{78.0} \\

\textbf{GaMMA-14B (ours)} & 14B
& \underline{78.6} & \underline{88.6} & \textbf{90.0} & \textbf{85.7} & \textbf{78.3} & \textbf{79.5} & \textbf{84.2} & 82.1 & \textbf{82.6} & \textbf{77.6} & \textbf{77.8} & \textbf{81.5} & 77.2 & \textbf{79.0} \\

    \bottomrule
  \end{NiceTabular}}
\end{table*}

\begin{table*}[t]
\caption{
  Experiment results on the MusicBench-Global. The benchmark assesses models on Global attributes (primary genre (PG), secondary genre (SG), mood), Fine-Grained perception (solo instrument (SI), lead instrument (LI), main accompanying instrument (MAI), gender, timbre, artist, language (Lang.), contains musical instruments (CMI), key, BPM, melody, rhythm), and Reasoning \& Association (R \& A) (scenario reasoning (SR), content association (CA)). 
}
  \label{tab:music-g}
  \centering
  \setlength{\tabcolsep}{0.7mm}
  \scalebox{0.75}{\begin{NiceTabular}{l*{19}{c}}
    \CodeBefore
      \columncolor{teal!7}{20} 
    \Body
    
    \toprule
    {} & \multirow{2}{*}{\textbf{Param.}} & \multicolumn{3}{c}{\textbf{Global}}  & \multicolumn{12}{c}{\textbf{Fine-Grained}} & \multicolumn{2}{c}{\textbf{R \& A}} & \multirow{2}{*}{\textbf{Acc.}} \\
    \cmidrule[0.5pt](rl){3-5}
    \cmidrule[0.5pt](rl){6-17}
    \cmidrule[0.5pt](rl){18-19}
    
    & &PG &SG &Mood &SI &LI &MAI &Gender &Timbre &Artist  &Lang. &CMI &Key &BPM &Melody &Rhythm &SR &CA &{} \\
    \midrule 
\textcolor{gray}{Gemini-2.0-Pro} & \textcolor{gray}{-}  & \textcolor{gray}{94.3} & \textcolor{gray}{91.0} & \textcolor{gray}{75.8} & \textcolor{gray}{76.6} & \textcolor{gray}{67.1} & \textcolor{gray}{70.0} & \textcolor{gray}{74.0} & \textcolor{gray}{58.6} & \textcolor{gray}{73.3} & \textcolor{gray}{83.0} & \textcolor{gray}{80.0} & \textcolor{gray}{28.5} & \textcolor{gray}{53.5} & \textcolor{gray}{43.2} & \textcolor{gray}{74.5} & \textcolor{gray}{69.5} & \textcolor{gray}{79.0} & \textcolor{gray}{70.3} \\
\textcolor{gray}{Gemini-2.5 Pro} & \textcolor{gray}{-} & \textcolor{gray}{92.2} & \textcolor{gray}{94.5} & \textcolor{gray}{87.4} & \textcolor{gray}{93.8} & \textcolor{gray}{81.7} & \textcolor{gray}{85.7} & \textcolor{gray}{90.3} & \textcolor{gray}{84.6} & \textcolor{gray}{89.7} & \textcolor{gray}{99.0} & \textcolor{gray}{70.0} & \textcolor{gray}{41.5} & \textcolor{gray}{49.3} & \textcolor{gray}{53.9} & \textcolor{gray}{77.9} & \textcolor{gray}{76.2} & \textcolor{gray}{84.5} & \textcolor{gray}{76.7} \\

\textcolor{gray}{Gemini-3.0 Pro} & \textcolor{gray}{-} & \textcolor{gray}{96.0} & \textcolor{gray}{96.0} & \textcolor{gray}{87.8} & \textcolor{gray}{95.7} & \textcolor{gray}{78.1} & \textcolor{gray}{78.6} & \textcolor{gray}{86.0} & \textcolor{gray}{80.0} & \textcolor{gray}{93.3} & \textcolor{gray}{98.5} & \textcolor{gray}{50.0} & \textcolor{gray}{58.0} & \textcolor{gray}{52.5} & \textcolor{gray}{56.4} & \textcolor{gray}{78.5} & \textcolor{gray}{81.0} & \textcolor{gray}{80.0} & \textcolor{gray}{80.4} \\

Mu-LLaMA~\cite{liu2024music} & 7B & 27.3 & 15.0 & 16.4 & 23.4 & 32.9 & 28.6 & 43.0 & 27.1 & 16.7 & 21.0 & 50.0 & 21.0 & 32.0 & 22.8 & 27.0 & 14.0 & 4.0 & 22.7 \\
M2UGen~\cite{liu2023m} & 7B & 20.5 & 24.0 & 26.4 & 31.9 & 16.4 & 25.7 & 26.0 & 38.6 & 46.7 & 25.0 & 30.0 & 21.5 & 29.5 & 24.9 & 25.0 & 33.0 & 32.0 & 26.0 \\
SALMONN~\cite{tang2023salmonn} & 13B & 24.5 & 18.0 & 29.0 & 29.8 & 28.8 & 31.4 & 23.0 & 27.1 & 33.3 & 22.5 & 30.0 & 36.0 & 27.0 & 24.9 & 26.0 & 21.0 & 36.0 & 26.8 \\
QwenAudio~\cite{chu2023qwen}& 7B & 67.5 & 45.0 & 70.6 & 51.1 & 56.2 & 30.0 & 45.0 & 54.3 & 26.7 & 71.0 & 60.0 & 29.5 & 35.0 & 29.1 & 60.0 & 38.5 & 35.0 & 52.0 \\
Qwen2Audio~\cite{chu2024qwen2}& 7B
& 74.3 & 76.0 & 75.2 & 68.1 & 60.3 & 42.9 & 62.0 & 42.9 & 80.0 & 77.0 & \underline{80.0} & 20.5 & 31.0 & 29.5 & 52.5 & 37.5 & 40.0 & 56.0 \\

Kimi-Audio~\cite{ding2025kimi}& 8B
& 88.8 & 83.0 & 78.2 & 70.2 & 78.1 & 70.0 & 86.0 & 72.9 & \underline{80.0} & 91.5 & \underline{80.0} & 26.0 & 40.0 & 42.7 & 48.0 & 61.5 & \textbf{80.0} & 67.6 \\

Audio-Flamingo3~\cite{goel2025audio} & 8B
& 71.0 & 79.6 & 83.3 & 77.8 & 61.1 & 60.6 & 82.2 & 53.2 & 58.6 & 96.9 & 50.0 & 80.0 & \underline{60.3} & 42.0 & 72.9 & 66.7 & 72.1 & 72.8 \\

Qwen2.5-Omni~\cite{xu2025qwen2} & 8B
& 94.0 & \textbf{95.0} & 83.0 & \underline{89.0} & \textbf{96.0} & \textbf{90.0} & \textbf{95.0} & \underline{77.0} & 77.0 & \underline{98.0} & \textbf{90.0} & 65.0 & 51.0 & 44.0 & 74.0 & 71.0 & \underline{79.0} & 78.0 \\

\midrule

\textbf{GaMMA-8B (ours)} & 8B
& \textbf{96.8} & \underline{93.0} & \textbf{84.0} & 87.2 & 89.0 & \underline{88.6} & \underline{90.0} & \textbf{88.6} & \textbf{83.3} & \textbf{99.5} & \underline{80.0} & \underline{87.5} & \textbf{61.0} & \underline{50.6} & \underline{84.0} & \underline{74.0} & 78.0 & \textbf{82.6} \\

\textbf{GaMMA-14B (ours)}& 14B
& \underline{95.5} & \underline{93.0} & \underline{83.4} & \textbf{89.4} & \underline{90.4} & 87.1 & \underline{90.0} & \textbf{88.6} & 76.7 & 97.5 & \underline{80.0} & \textbf{89.5} & 48.5 & \textbf{51.5} & \textbf{85.0} & \textbf{93.0} & \underline{79.0} & \underline{81.3} \\

\bottomrule
\end{NiceTabular}}
\end{table*}

\noindent\textbf{Experiments on MusicBench.} We evaluate various LAMMs on MusicBench, with results summarized in Tab.~\ref{tab:music-g}-\ref{musicbench}. As shown in the table, GaMMA-8B achieves the highest overall accuracy at \(82.6\%\), surpassing all baselines, including Gemini-3.0 Pro (\(80.4\%\)). A key highlight of GaMMA lies in its superior performance on temporal understanding tasks. As shown in Table~\ref{tab:MusicBench-Temporal}, GaMMA-14B achieves \(95.5\%\) on lyrics, \(86.5\%\)  on structure, and \(79.3\%\)  overall—ranking outperforms Gemini-3.0 Pro \(75.2\%\) while using fewer parameters. Notably, GaMMA-8B also achieves state-of-the-art performance among models of similar size (\(76.2\%\) \vs \(65.5\%\)), underscoring its strong performance in the temporal domain. These results underscore the ability of GaMMA to effectively capture complex temporal dynamics in music with a unified parameter set.

\begin{table}[h]
\renewcommand{\arraystretch}{1.2}
\setlength{\tabcolsep}{12pt} 
\caption{Experiment results on MusicBench-Temporal.}
\vspace{0.05in}
\scalebox{0.8}{\begin{NiceTabular}{ccccccc}
    \CodeBefore
      \columncolor{teal!7}{7} 
    \Body
    
    \toprule
    {} & \textbf{Chords} & \textbf{Structure} & \textbf{Lyrics} & \textbf{Inst.} & \textbf{Vocal} & \textbf{Acc.} \\ 
    
    \midrule
    \textcolor{gray}{Gemini-2.5 Pro} &\textcolor{gray}{46.5} &\textcolor{gray}{91.5} & \textcolor{gray}{97.0} & \textcolor{gray}{69.2} & \textcolor{gray}{64.8} & \textcolor{gray}{74.6}   \\
    \textcolor{gray}{Gemini-3.0 Pro} &\textcolor{gray}{53.0} &\textcolor{gray}{88.0} & \textcolor{gray}{89.5} & \textcolor{gray}{74.5} & \textcolor{gray}{70.4} & \textcolor{gray}{75.2}   \\
    Mu-LLaMA~\cite{liu2024music} & 15.5 & 28.0 & 3.5 & 29.2 & 34.2 & 22.0 \\
    QwenAudio~\cite{chu2023qwen} & 25.0 & 35.0 & 29.5 & 51.8 & 33.2 & 34.9 \\
    Qwen2Audio~\cite{chu2024qwen2} & 28.5 & 38.0 & 57.0 & 50.3 & 32.2 & 41.2 \\
    Flamingo 2~\cite{alayrac2022flamingo} & 27.0 & 32.5 & 15.5 & 26.6 & 35.7 & 27.5 \\
    Kimi-Audio~\cite{ding2025kimi} & 27.5 & 49.5 & 75.0 & 61.8 & 37.2 & 50.3 \\
    Audio-Flamingo3~\cite{goel2025audio} & 46.0 & 57.0 & 35.0 & 49.8 & 44.7 & 46.5 \\
    Qwen2.5-Omni~\cite{xu2025qwen2} & 45.0 & 57.0 & 86.0 & 78.4 & \underline{61.3} & 65.5 \\
    \midrule
    \textbf{GaMMA-8B (ours)} & \underline{60.0} & \textbf{92.0} & \underline{90.5} & \textbf{82.4} & 55.8 & \underline{76.2} \\
    \textbf{GaMMA-14B (ours)} & \textbf{75.0} & \underline{86.5} & \textbf{95.5} & \underline{70.0} & \textbf{68.9} & \textbf{79.3} \\
    \bottomrule 
\end{NiceTabular}}
\label{tab:MusicBench-Temporal}
\end{table}

\noindent\textbf{Experiments on MMAU.} Beyond music understanding, we further evaluate the generalization ability of GaMMA on the MMAU, which covers a broad range of audio tasks, including sound, music, and speech. As shown in Table~\ref{tab:mmau_results}, GaMMA-14B achieves a competitive average accuracy of \(75.8\%\), on par with the strongest baseline Gemini-3.0 Pro (\(78.4\%\)), and notably outperforms all models in the music category with a remarkable \(74.0\%\). It also delivers strong performance on speech tasks (\(73.6\%\)). This highlights the potential of GaMMA as a unified solution for broad-spectrum audio-language understanding beyond music-centric tasks. 

\begin{table}[h]
\renewcommand{\arraystretch}{1.2}
\setlength{\tabcolsep}{14pt} 
\caption{Results on MMAU-mini-v05.15.25.}
\vspace{0.05in}
\scalebox{0.8}{\begin{NiceTabular}{l cccc}
    \CodeBefore
      \columncolor{teal!7}{5}
    \Body
  
    \toprule
    {}  & \textbf{Sound} & \textbf{Music} & \textbf{Speech} & \textbf{Avg.} \\ 
    \midrule
    \textcolor{gray}{Gemini-2.5 Pro}   & \textcolor{gray}{75.1}  & \textcolor{gray}{68.3} & \textcolor{gray}{71.5} & \textcolor{gray}{71.6} \\
    \textcolor{gray}{Gemini-3.0 Pro}   & \textcolor{gray}{79.6}  & \textcolor{gray}{72.2} & \textcolor{gray}{83.5} & \textcolor{gray}{78.4} \\
    \textcolor{gray}{GPT-4o Audio}   & \textcolor{gray}{64.6}  & \textcolor{gray}{56.3} & \textcolor{gray}{66.7} & \textcolor{gray}{60.8} \\
    SALMONN~\cite{tang2023salmonn} &41.1 &37.1 &26.4 &34.9 \\
Qwen2-Audio~\cite{chu2024qwen2} & 67.3 & 56.3 & 55.3 & 59.6 \\

Kimi-Audio~\cite{ding2025kimi}   & 75.7 & 66.8 & 62.2 & 68.2 \\

Audio-Flamingo3~\cite{goel2025audio} & \underline{79.6} & \textbf{74.0} & 66.4 & \underline{73.3} \\

Qwen2.5-Omni~\cite{xu2025qwen2}   & 78.1 & 65.9 & \underline{70.6} & 71.5 \\

\midrule

\textbf{GaMMA-8B (ours)}  & 76.3 & \underline{73.7} & 64.0 & 71.3 \\

\textbf{GaMMA-14B (ours)} & \textbf{79.9} & \textbf{74.0} & \textbf{73.6} & \textbf{75.8} \\

    \bottomrule
  \end{NiceTabular}}
  \label{tab:mmau_results}
\end{table}


\subsection{Expert Study}

\begin{table*}[htbp]
\caption{
\textbf{Expert study results.} We compare GaMMA with Gemini-2.5 Pro and Gemini-3.0 Pro across general dimensions (vocal, language (Lang.), rhythm, genre, BPM, mood, theme, scenario reasoning (SR), content association (CA), highlight description, instrument (Inst.), and melody) and temporal dimensions (first verse–chorus progression (FVCP), structural segmentation (Structure), instrument (Inst.), rhythm, melody, key and mode (K\&M), and chord progression (Chord)).
}
  \label{tab:expert_main}
  \centering
  \setlength{\tabcolsep}{0.7mm}
  \scalebox{0.66}{\begin{tabular}{l*{19}{c}c}
    \toprule
{} & \multicolumn{11}{c}{\textbf{Global}}  & \multicolumn{5}{c}{\textbf{Temporal}}   \\
\cmidrule[0.5pt](rl){2-13}
\cmidrule[0.5pt](rl){14-20}

        &Vocal &Lang.  &Genre &Rhythm &BPM  &Mood &Theme &SR &CA &Highlight &Inst. &Melody &FVCP & Structure &Inst. &Rhythm & Melody &K \& M & Chord  \\
    \midrule

Gemini-2.5 Pro & 91.0  & 99.0  & 91.2 & \textbf{91.6} & 23.6 & 98.2 & 98.8 & 90.0 & 99.2 & 96.2 &\textbf{ 85.6} & 97.0 & 60.4 & 57.6 & 80.8 &79.4 &89.8 &11.6 & 0.0 \\
Gemini-3.0 Pro & 82.0 & 92.0 &87.0 &71.0 & 0.0 &91.0 &94.0 &87.0 &93.0 &94.0 &77.0 &89.0 &6.0 &1.0 &56.0 &66.0 &68.0 &16.8 &0.0  \\

\midrule

\textbf{GaMMA-14B (ours)} &\textbf{95.8} & \textbf{99.0} &\textbf{92.6} &91.0 &\textbf{95.8} &\textbf{99.6} &\textbf{99.0} &\textbf{100.0} & \textbf{99.6} &\textbf{99.2} &85.4 &\textbf{99.6} &\textbf{88.4} &\textbf{91.4} &\textbf{85.4} &\textbf{95.5} &\textbf{99.4} &\textbf{82.6} &\textbf{72.5} \\
\bottomrule
\end{tabular}}
\vspace{10pt}

\end{table*}
While MusicBench provides a comprehensive evaluation of discriminative capabilities via multiple-choice questions, it is equally critical to assess the generative proficiency of model in open-ended natural language tasks. To this end, we conducted a rigorous human evaluation study.

\textbf{Experimental Setup.}
We recruited 10 music experts, including professional music practitioners and graduate students with formal training in music. Besides, we curated a test set of 500 music tracks, balancing musical genres and languages to ensure diversity.

\textbf{Evaluation Protocol and Metrics.}
We evaluate model outputs along global understanding and temporal understanding. The fine-grained criteria under each category follow the definitions in MusicBench for consistency. Evaluators rated each model output using a 4-point Likert scale from 1 to 4. As the primary metric, we report the \emph{Usability Rate}, defined as the percentage of samples with scores $\ge 3$ for each dimension. Details on the expert study are provided in Sec.~\ref{sec:supp_expert}.

\textbf{Expert Study Results.} Table~\ref{tab:expert_main} reports the human evaluation results measured by usability rate. Overall, GaMMA consistently outperforms both Gemini-2.5 Pro and Gemini-3.0 Pro across global and temporal dimensions. The most prominent gains come from temporal reasoning, GaMMA achieves high usability on structure-related skills (\eg, verse-chorus progression and structural analysis) and shows non-trivial capability on temporal K \& M and chord progression, where the Gemini baselines remain weak or fail entirely. We present a qualitative visualization comparison between GaMMA and existing SoTA models in Fig.~\ref{fig:com_vis}. Compared with Kimi-Audio~\cite{ding2025kimi}, Qwen2.5-Omni~\cite{xu2025qwen2} and Gemini-3.0 Pro, GaMMA achieves more precise temporal localization and a much more coherent analysis of the buildup.

\subsection{Ablation} 
\label{ablation_Sec}

\begin{minipage}[htbp]{0.3\textwidth}
\makeatletter\def\@captype{table}
\centering
\caption{Ablation on music-lyrics data.}
\vspace{0.05in}
   \scalebox{0.6}{\begin{tabular}{lccc}
     \toprule
     &{} & \textbf{MuCho.} & \textbf{MusicBench-G}\\
     \midrule

     & \(w/o\) lyrics data & 60.8 & 68.1\\
     & \(w.\) lyrics data  & \textbf{63.6} & \textbf{69.8} \\
     \bottomrule
   \end{tabular}}
\label{tab:audio-language}
\end{minipage}
\begin{minipage}[htbp]{0.33\textwidth}
\makeatletter\def\@captype{table}
\centering
   \caption{Ablation study on encoder architecture and fusion methods.}
   \vspace{0.05in}
   \scalebox{0.5}{\begin{tabular}{lccc}
     \toprule
     \textbf{Architecture} & \textbf{MuCho.} & \textbf{MusicBench-G} & \textbf{Harmonix} \\
     \midrule
     Single Encoder        & 32.1 & 43.5 & 57.1 \\
     Dual Encoder + MLP    & 60.7 & 64.4 & 50.7 \\
     DFN    & \textbf{67.6} & \textbf{74.8} & \textbf{57.1} \\
     \bottomrule
   \end{tabular}}
   \label{tab:ablation_architecture}
\end{minipage}
\begin{minipage}[htbp]{0.37\textwidth}
\makeatletter\def\@captype{table}
\centering
   \caption{Ablation study on training strategy.}
\vspace{0.05in}
   \scalebox{0.5}{\begin{tabular}{lccc}
     \toprule
     {} & \textbf{MuCho.} & \textbf{MusicBench-G}  & \textbf{MusicBench-T}\\
     \midrule
     pretrain & 63.1 & 66.0  & 38.4\\
     + SFT & 67.9 & 79.3 & 61.2\\
     + GRPO  & \textbf{71.0}  & \textbf{80.6} & \textbf{63.1}\\
     \bottomrule
   \end{tabular}}
   \label{tab:ablation_grpo}
\end{minipage}

In this section, we present ablation studies to analyze the impact of different training settings and architectural components on the performance of GaMMA.

\noindent\textbf{Is Audio-Language Alignment Necessary Beyond the Audio-Caption Alignment?}
As shown in Table~\ref{tab:audio-language}, leveraging lyrics data brings benefit to downstream performance, yielding improvements of \(+2.8\%\) on MuChoMusic and \(+1.7\%\) on MusicBench. This finding suggests that audio-language alignment serves as an important complement to audio-caption alignment, enriching music comprehension of GaMMA across training stages.

\noindent\textbf{Single-encoder architecture struggles with joint processing of time-series and non-time-series data.} We leverage Harmonix~\cite{nieto2019harmonix}, a benchmark designed to assess the music structure analysis abilities of models in this part of the experiments. Additional details about Harmonix are provided in Sec.~\ref{sec:supp_harmonix}. In earlier experiments, we observed that effective performance on temporal tasks could only be achieved when the parameters of the audio encoder were unfrozen. However, as shown in Table~\ref{tab:ablation_architecture}, jointly training on both temporal and non-temporal data with an unfrozen single encoder led to a significant performance drop on non-temporal benchmarks such as MuChoMusic and MusicBench. This indicates that a single encoder lacks the representational capacity to simultaneously model both temporal dynamics and high-level semantics.

\noindent\textbf{Effectiveness of DFN Fusion.} Motivated by the above finding, we adopt a dual-encoder architecture, using separate audio encoders specialized for time-series and non-time-series inputs, and explore different fusion strategies. Compared to simple MLP-based fusion, our proposed DFN more effectively integrates embeddings from the two experts. As shown in Table~\ref{tab:ablation_architecture}, DFN achieves significant improvements of \(+6.9\%\) on MuChoMusic and \(+10.4\%\) on MusicBench over the MLP baseline, demonstrating its superior ability to balance temporal and non-temporal understanding.

\noindent\textbf{Effectiveness of training strategy.}
As shown in Table~\ref{tab:ablation_grpo}, the base model pretrained on large-scale music-caption and music-lyric pairs already achieves a solid foundation across benchmarks (\eg, \(66.0\%\) on MusicBench-G), but performs poorly on temporal tasks (\(38.4\%\) on MusicBench-T), indicating limited temporal understanding.
Incorporating supervised fine-tuning (SFT) with instruction-tuned and temporally annotated data yields significant gains across all metrics, particularly on temporal understanding (\(+22.8\%\) on MusicBench-T).
We introduce GRPO in the last stage of training. Experiment results show that it further improves overall performance on MuChoMusic and MusicBench-T, demonstrating its effectiveness in refining multi-choice reasoning and boosting overall QA accuracy under reinforcement learning.

%% file: sec/5_con.tex
\section{Conclusion}
We introduce GaMMA, a large audio multimodal model designed to capture the full complexity of musical content. By combining DFN with a three-stage progressive training strategy, GaMMA jointly addresses time-series and non-time-series tasks within one set of parameters. To benchmark progress in this area, we also present MusicBench, the largest and most comprehensive human-curated benchmark for music LMMs. GaMMA achieves SoTA on both MusicBench and MuChoMusic, demonstrating its effectiveness in general music understanding. We believe GaMMA represents a significant step forward in enabling LMMs to interact more naturally with real-world environments, where music plays an integral role in human experience.

%% file: sec/X_appen.tex
\section{Details on the Training Strategy and Datasets}
\label{sec:supp_strategy}

\begin{figure*}[h]
  \centering
  \includegraphics[width=0.95\linewidth]{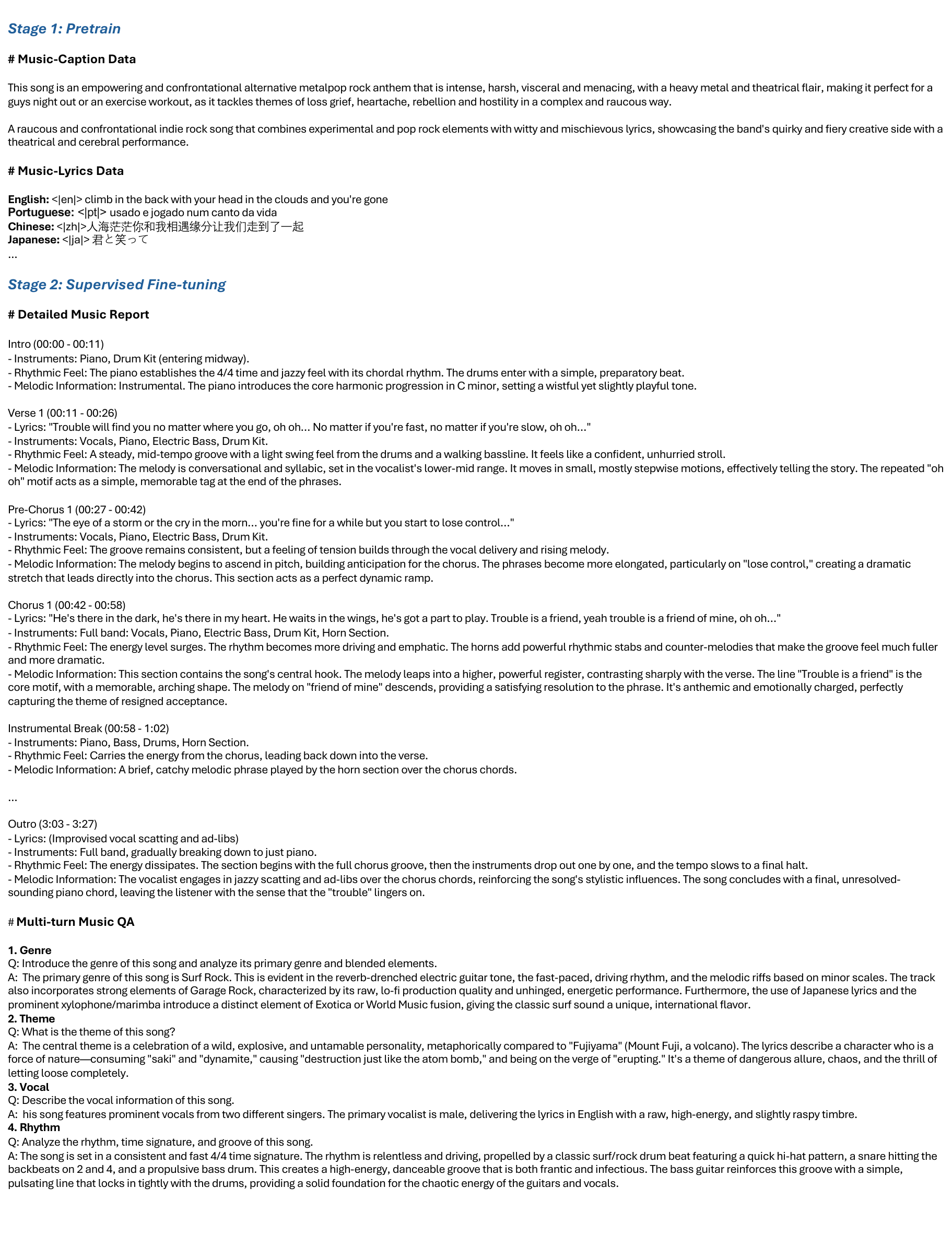}
  \caption{Data visualization across training stages.}
  \label{dataset_vis1}
\end{figure*}

\begin{figure*}[h]
  \centering
  \includegraphics[width=0.95\linewidth]{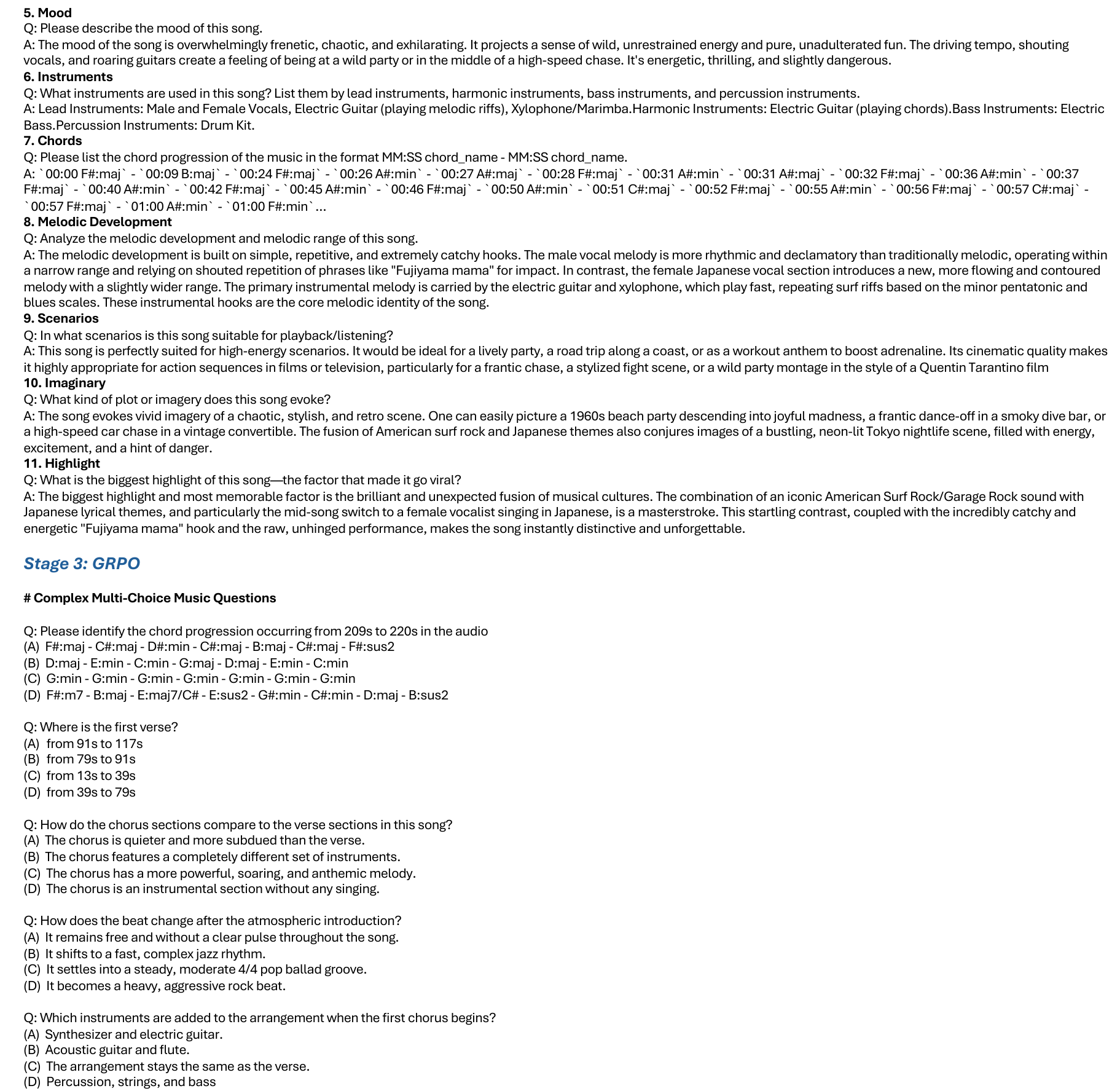}
  \caption{Data visualization across training stages (continue).}
  \label{dataset_vis2}
\end{figure*}

In this section, we detail the training strategy and datasets used for GaMMA. As summarized in Table~\ref{tab:settings}, we present the data scales and key hyperparameter settings. Fig.~\ref{dataset_vis1}-~\ref{dataset_vis2} provide visualizations of the datasets across different training stages.

\begin{table}[h]
  \caption{Training configuration for each stage. }
  \label{tab:settings}
  \centering
  \setlength{\tabcolsep}{2.0mm}
  \begin{tabular}{lccc}
    \toprule
    {} & Stage-1 & Stage-2 & Stage-3\\
    \midrule
    \textbf{AudioLength} & Max 60s & Max 300s & Max 120s\\
    \#Tokens & Max 1500 & Max 7500 & Max 3000 \\
    \#Samples & 5.3M & 1.6M &1.7K\\
    \textbf{Trainable} & Projector & Full Model & Full Model\\
    \textbf{Batch Size} & 256 & 64 & 16\\

    \bottomrule
  \end{tabular}
\end{table}

\noindent\textbf{\textit{Stage 1}: Multi-Modal Alignment Pretraining.} During the pretraining stage, we align music and language modalities using large-scale music-caption and music-lyric datasets. To align music with lyrics at a fine-grained level, we segment each music track into clips based on the timestamp of each lyric line, ensuring that each audio clip corresponds to a single lyric sentence. To enhance the multilingual capability of GaMMA, the music-lyric corpus includes lyrics in multiple languages. We represent the language of each lyric using special tokens from the vocabulary of Qwen~\cite{xu2025qwen2}, which are prepended to the lyric text during training.

\noindent\textbf{\textit{Stage 2}: Supervised Fine-tuning.}
To enhance both the temporal understanding and multi-turn music dialogue capabilities of GaMMA, we annotate approximately 60,000 music tracks with detailed structural reports. Each report contains fine-grained musical structure annotations along with precise start and end timestamps. We employ a highly qualified annotation team, whose members are graduates of renowned music academies and are well trained in music theory, to reannotate and verify these reports. For each structural unit, annotators provide detailed descriptions from four key perspectives: \textit{lyrics}, \textit{instrumentation}, \textit{rhythm}, and \textit{melodic information}. To further strengthen GaMMA’s ability to engage in multi-turn musical conversations, we construct a large-scale music dialogue dataset based on the annotated reports. The initial question templates used for dialogue generation are illustrated in Fig.~\ref{dataset_vis1} and Fig.~\ref{dataset_vis2}.

\noindent\textbf{\textit{Stage 3}: Reinforcement Learning.} In the final stage, we introduce reinforcement learning to further improve the performance of GaMMA on complex multiple-choice reasoning tasks, especially those requiring temporal and structural understanding. Specifically, we curated 1,723 high-quality multiple-choice questions that require reasoning over structural relationships (\eg, between sections, instruments, or motifs) using the pipeline in Sec.~\ref{sec:dataset_curation}. We present several QA samples used in the RL stage in Fig.~\ref{dataset_vis2}. Our data construction strategy ensures that questions are neither too easy nor too difficult, preventing the model from receiving trivial all-correct or all-incorrect signals during policy updates, and question re-synthesis further improves the model’s reasoning capabilities.

Note that all music tracks used during training are entirely different from those in evaluation benchmarks, ensuring a fair and unbiased evaluation.

\begin{figure}[t]
  \centering
  \includegraphics[width=\linewidth]{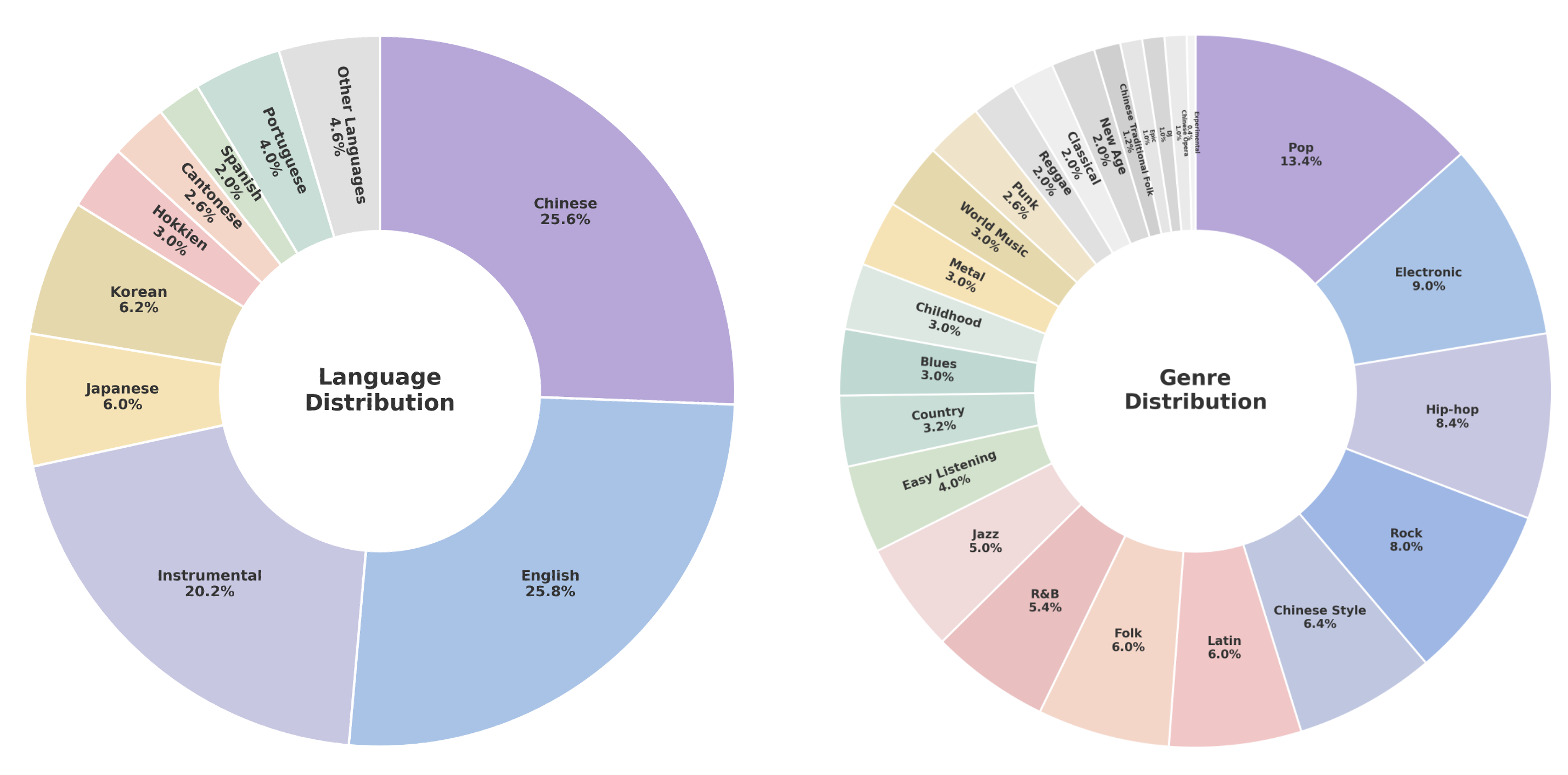}
  \caption{The distribution of the music for evaluation.}
  \label{fig:user_study_distribution}
\end{figure}

\section{Details on the Evaluation Metrics on Harmonix}
\label{sec:supp_harmonix}
To evaluate the music structure analysis ability of GaMMA, we adopt the segment-level F1-score as the core evaluation metric on the Harmonix. Following the setup used in MusicFM~\cite{won2024foundation}, and remap the original labels into seven unified segment categories: \textit{intro}, \textit{verse}, \textit{chorus}, \textit{bridge}, \textit{outro}, \textit{silence}, and \textit{instrumental}; and use the same test split. To align the predicted segments with the ground truth, we employ the Hungarian matching algorithm with an Intersection-over-Union (IoU) threshold of 0.5.

\section{Details on MusicBench}
\label{sec:supp_musicbench}

\begin{wrapfigure}{r}{0.6\columnwidth}
    \vspace{-40pt}
    \centering
\includegraphics[width=\columnwidth]{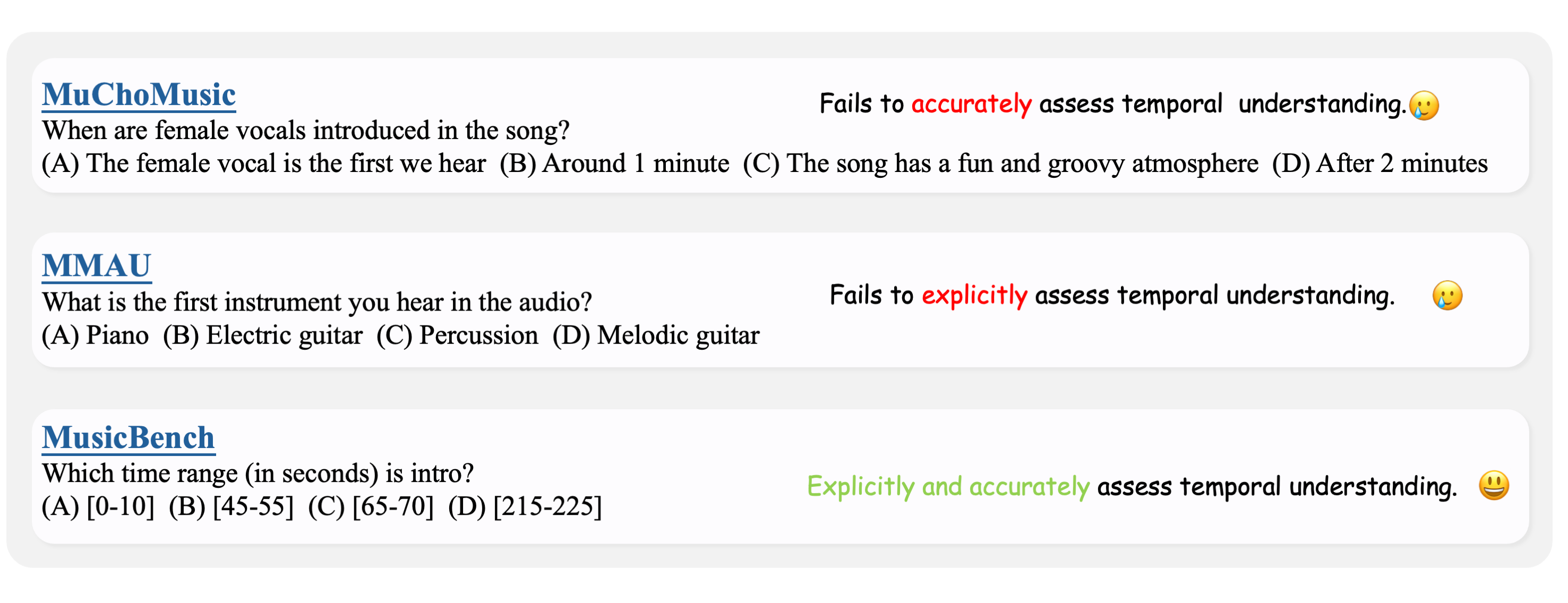}
  \caption{Comparison on temporal evaluation.}
  \label{fig:dataset_comparison}
    \vspace{-5pt}
\end{wrapfigure}

We provide additional details about MusicBench in this section. Table~\ref{tab:musicbench_details} illustrates several representative questions sampled from various subcategories within the benchmark. MusicBench offers a detailed and nuanced analysis of musical content, enabling comprehensive evaluation across a wide range of musical attributes. 

Notably, MusicBench is the \textbf{first benchmark} specifically designed to \textit{explicitly and comprehensively assess the temporal understanding} capabilities of music-language models. As shown in Fig.~\ref{fig:dataset_comparison}, both MuChoMusic and MMAU lack this level of evaluation. In contrast, MusicBench assesses temporal ability across lyrics, chords, structure, vocals, and instrumentation, offering a thorough and fine-grained evaluation of time-series music tasks.

\section{Expert Study Setting}
\label{sec:supp_expert}
Unlike other audio tasks, many characteristics of music cannot be easily evaluated through standard benchmarks. Certain dimensions that require specialized musical knowledge, such as key, mode, and the energy variations between verse and chorus, can only be accurately assessed by professionals with formal music training. To establish a reasonable framework for evaluating model capabilities, we designed a comprehensive evaluation protocol centered on the music audio itself. In this chapter, we provide a detailed description of our expert study, including the full evaluation methodology as well as the distribution of test music in terms of language and genre.

Fig.~\ref{fig:user_study_distribution} illustrates the distribution of languages and types of audio used in the expert study. English and Chinese are the most represented languages, followed by instrumental tracks. Additional languages, including Korean, Japanese, Hokkien, Cantonese, Spanish, and Portuguese, are also included, ensuring a truly multilingual evaluation. Regarding genre, the test set encompasses a broad spectrum of musical styles. Pop is the most common genre, followed by Electronic, Hip-hop, and Rock. The dataset also includes Chinese Style, Latin, Folk, R\&B, Jazz, Easy Listening, Country, Blues, Childhood, Metal, World Music, Punk, Classical, Classic Vocal, New Age, and other genres. This wide-ranging coverage supports a thorough assessment of model performance across diverse musical contexts.

To systematically assess model performance beyond surface-level correctness, we design a set of fine-grained evaluation rubrics as shown in Tab.~\ref{tab:rubric_1}-~\ref{tab:rubrics_2} that capture both factual accuracy and musical adequacy across multiple dimensions of a song. All rubrics adopt a 4–point ordinal scale, where a score of 4 corresponds to a fully correct, musically appropriate, and often optimal response; 3 indicates an acceptable but sub‑optimal answer; 2 denotes partially correct outputs with noticeable errors or omissions; and 1 is assigned when the response is essentially incorrect or musically irrelevant. This common structure allows us to compare behaviors across heterogeneous tasks while retaining task-specific nuance. 

\section{Qualitative Results}

\begin{figure*}[h]
  \centering
  \includegraphics[width=0.95\linewidth]{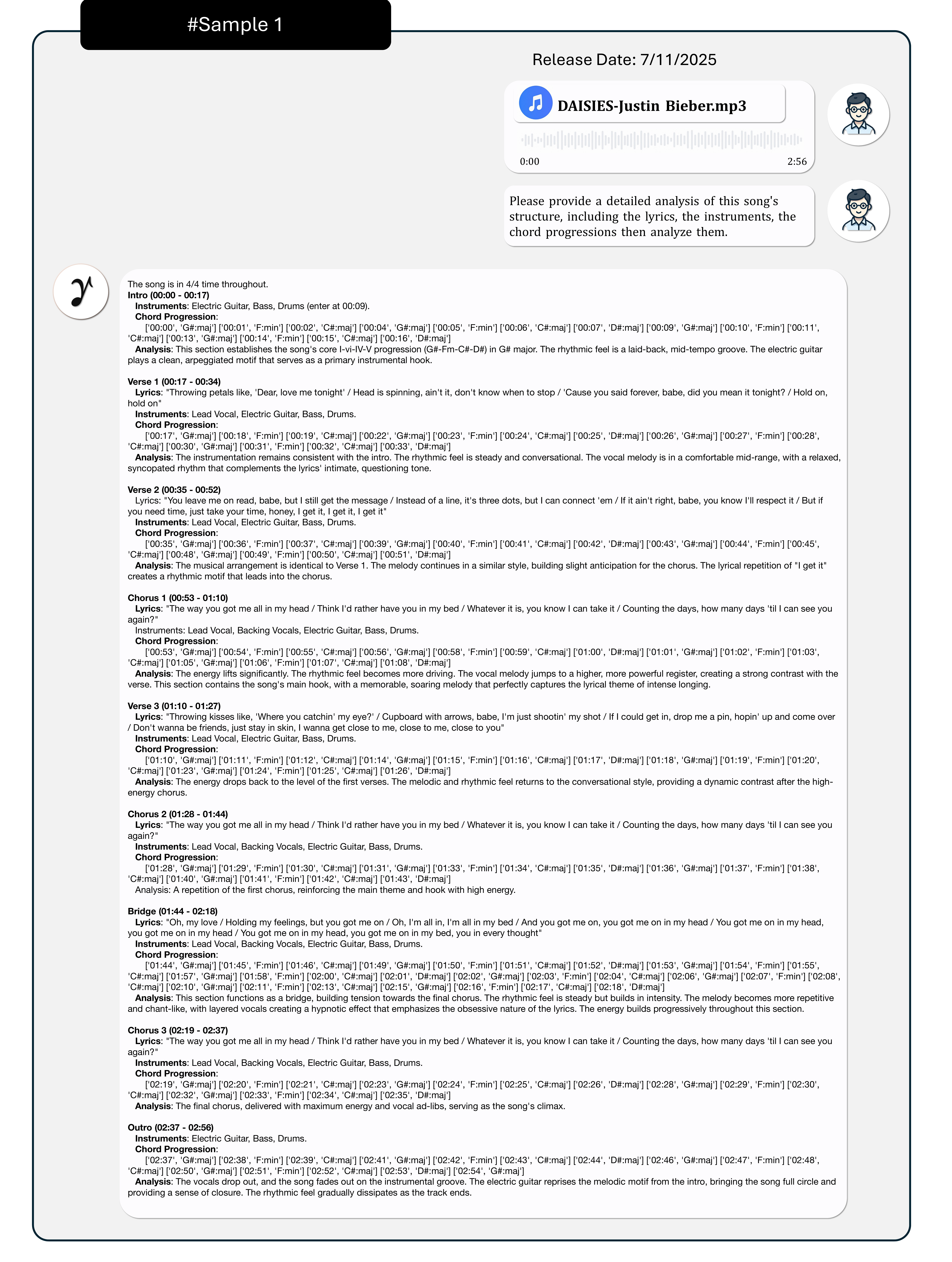}
  \caption{Conversation samples from GaMMA (1).}
  \label{sample1}
\end{figure*}

\begin{figure*}[h]
  \centering
  \includegraphics[width=0.95\linewidth]{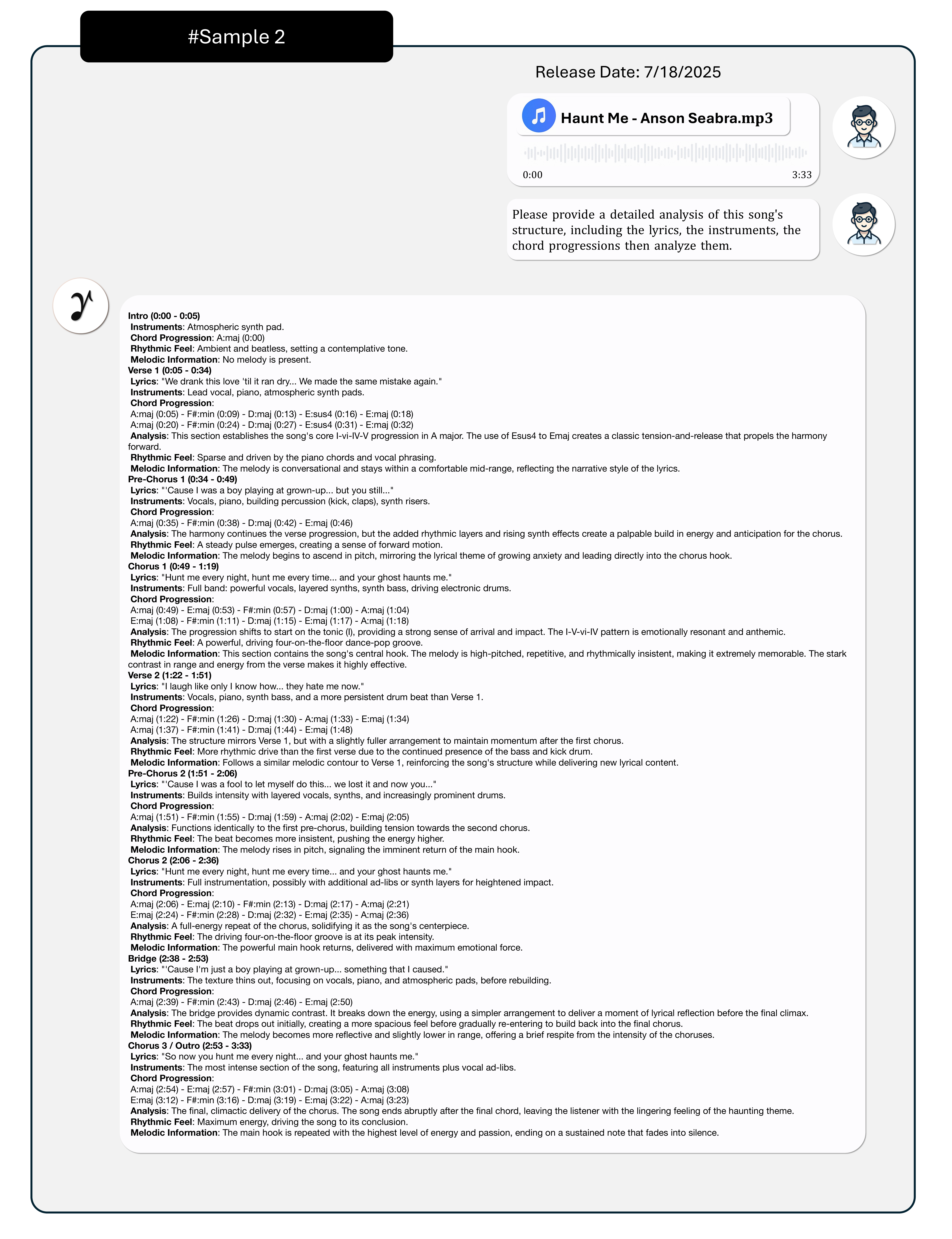}
  \caption{Conversation samples from GaMMA (2).}
  \label{sampl2}
\end{figure*}

\begin{figure*}[h]
  \centering
  \includegraphics[width=0.95\linewidth]{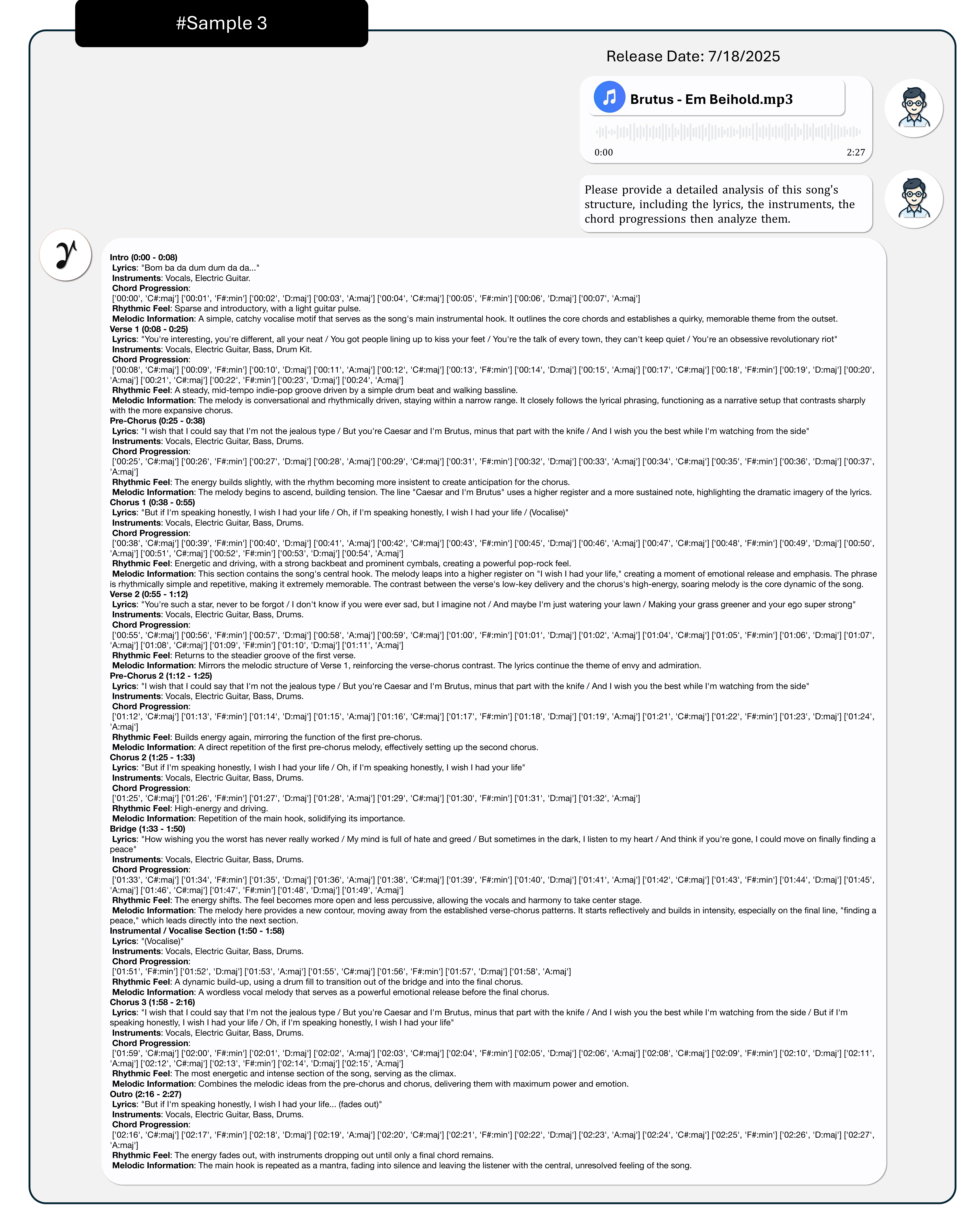}
  \caption{Conversation samples from GaMMA (3).}
  \label{sample3}
\end{figure*}

\begin{figure*}[h]
  \centering
  \includegraphics[width=0.95\linewidth]{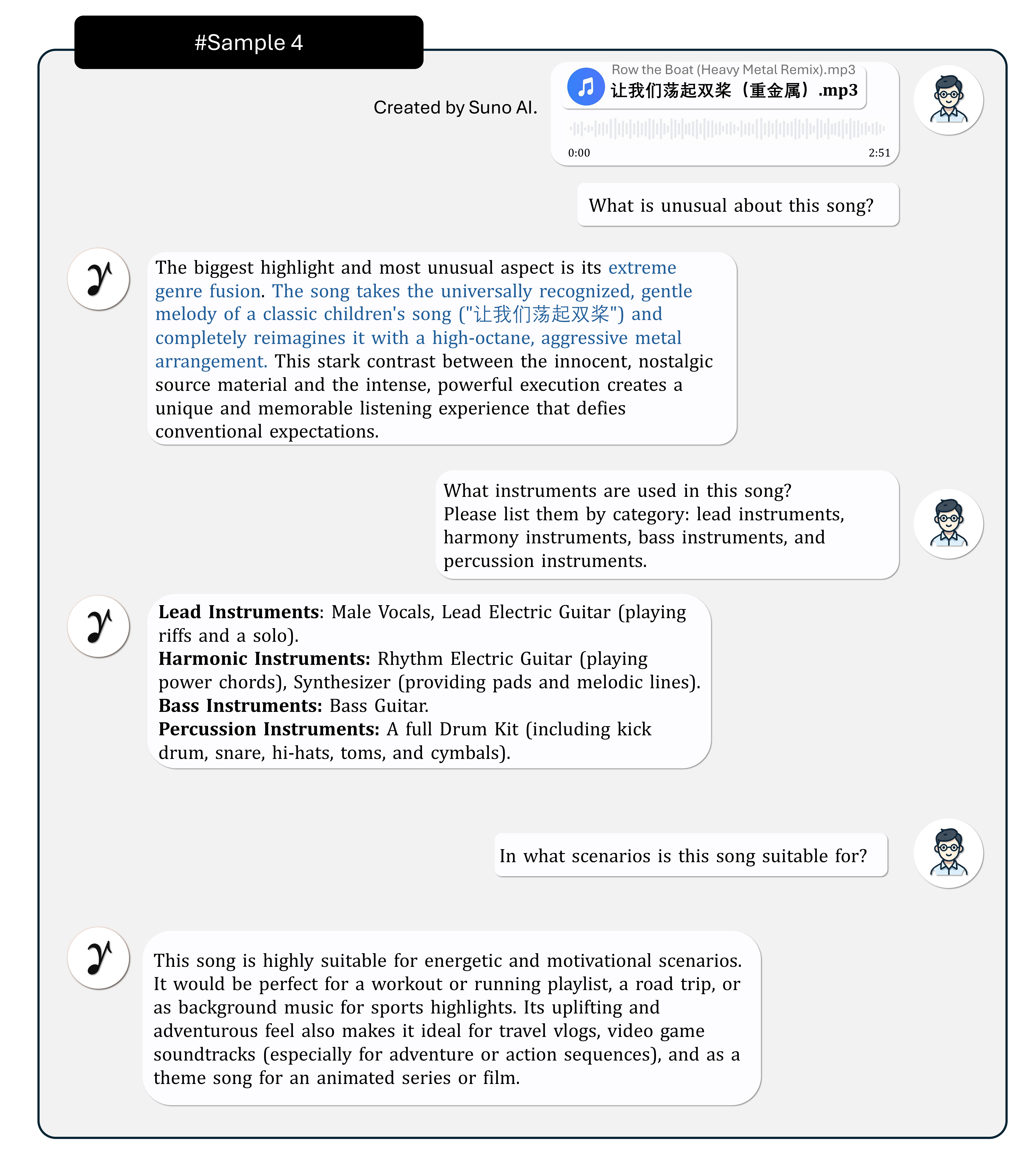}
  \caption{Conversation samples from GaMMA (4).}
  \label{sample4}
\end{figure*}

\begin{figure*}[h]
  \centering
  \includegraphics[width=0.95\linewidth]{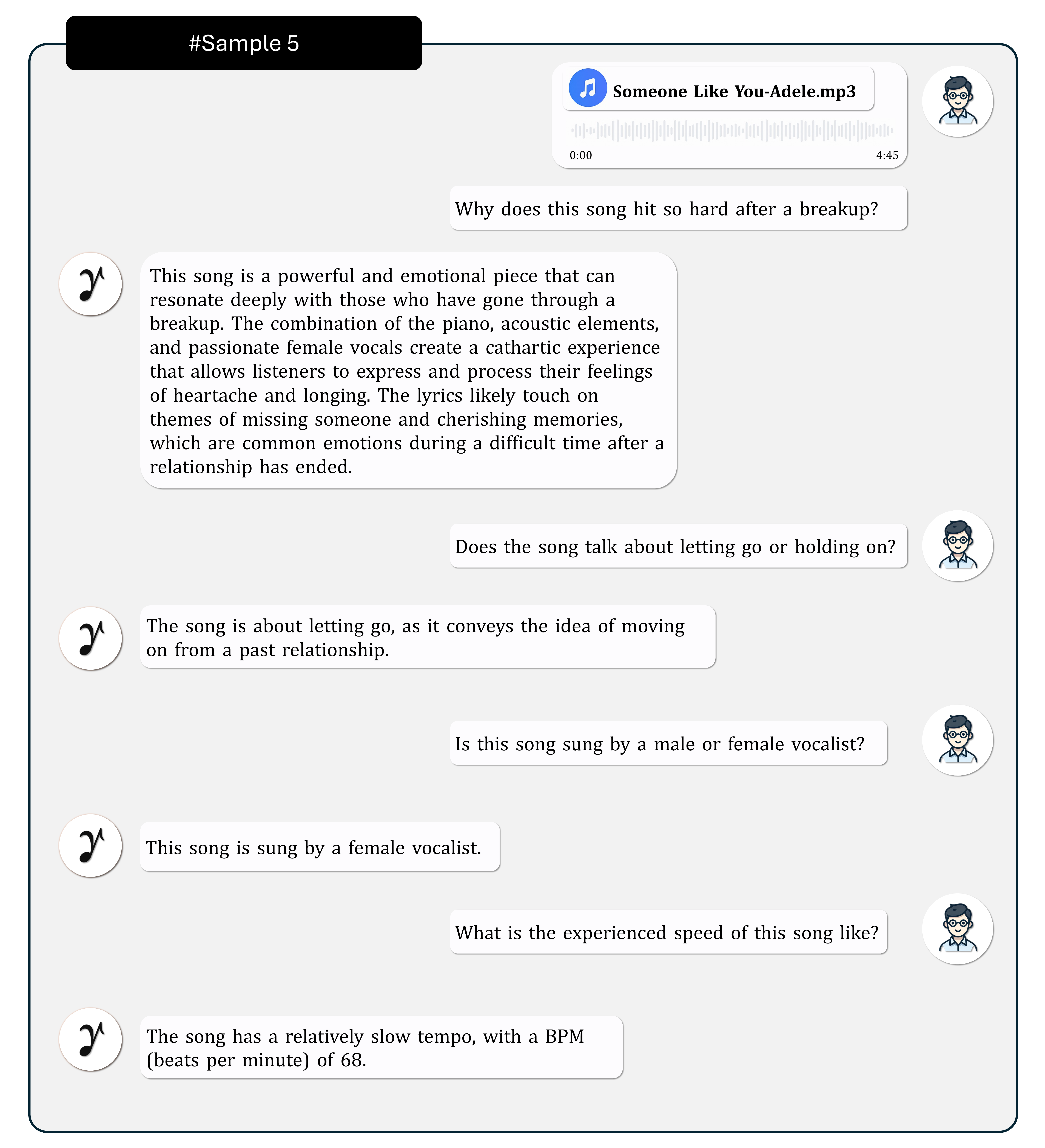}
  \caption{Conversation samples from GaMMA (5).}
  \label{sample5}
\end{figure*}

\begin{figure*}[h]
  \centering
  \includegraphics[width=0.95\linewidth]{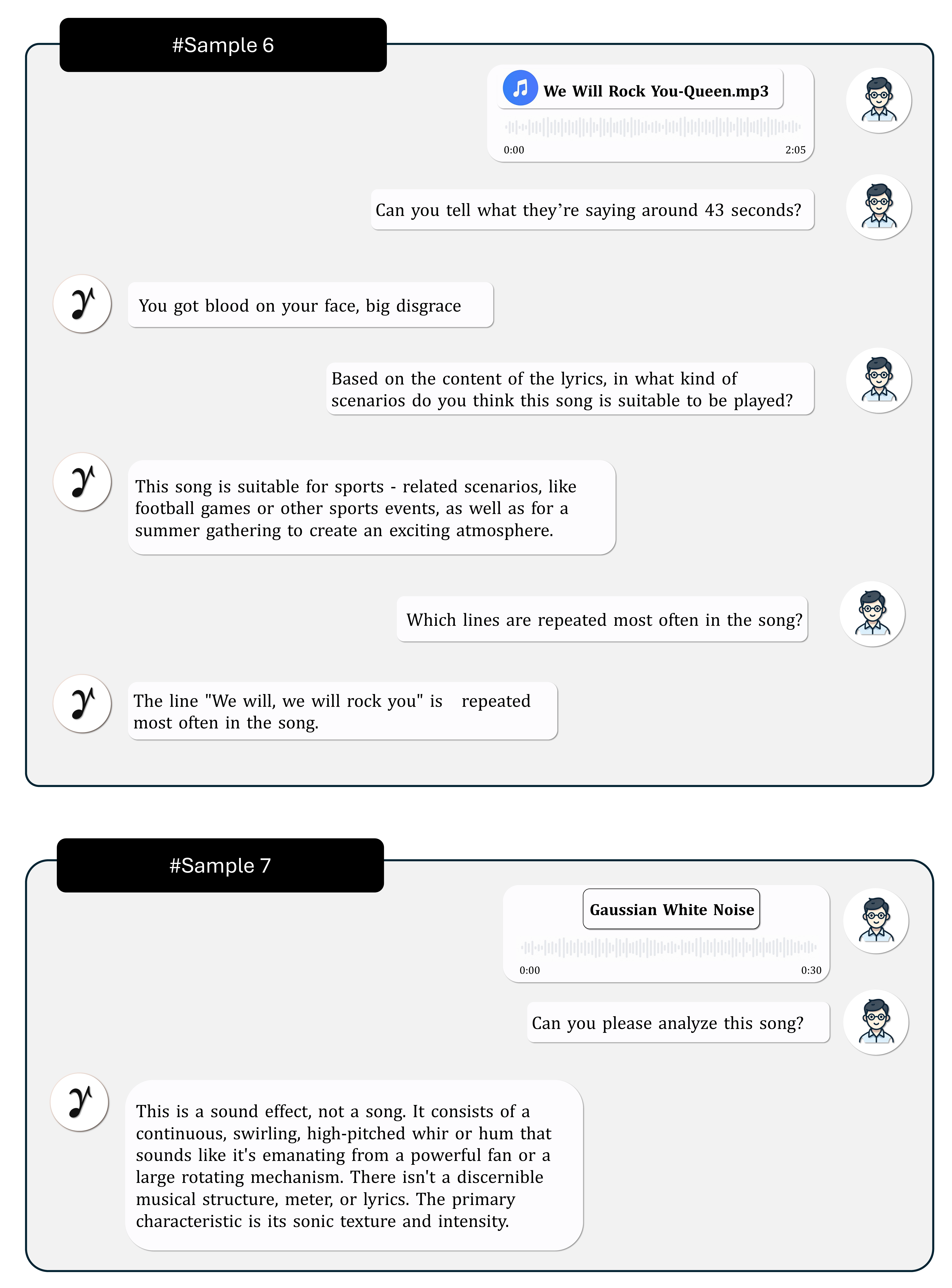}
  \caption{Conversation samples from GaMMA (6).}
  \label{sample6}
\end{figure*}

\begin{figure*}[h]
  \centering
  \includegraphics[width=0.95\linewidth]{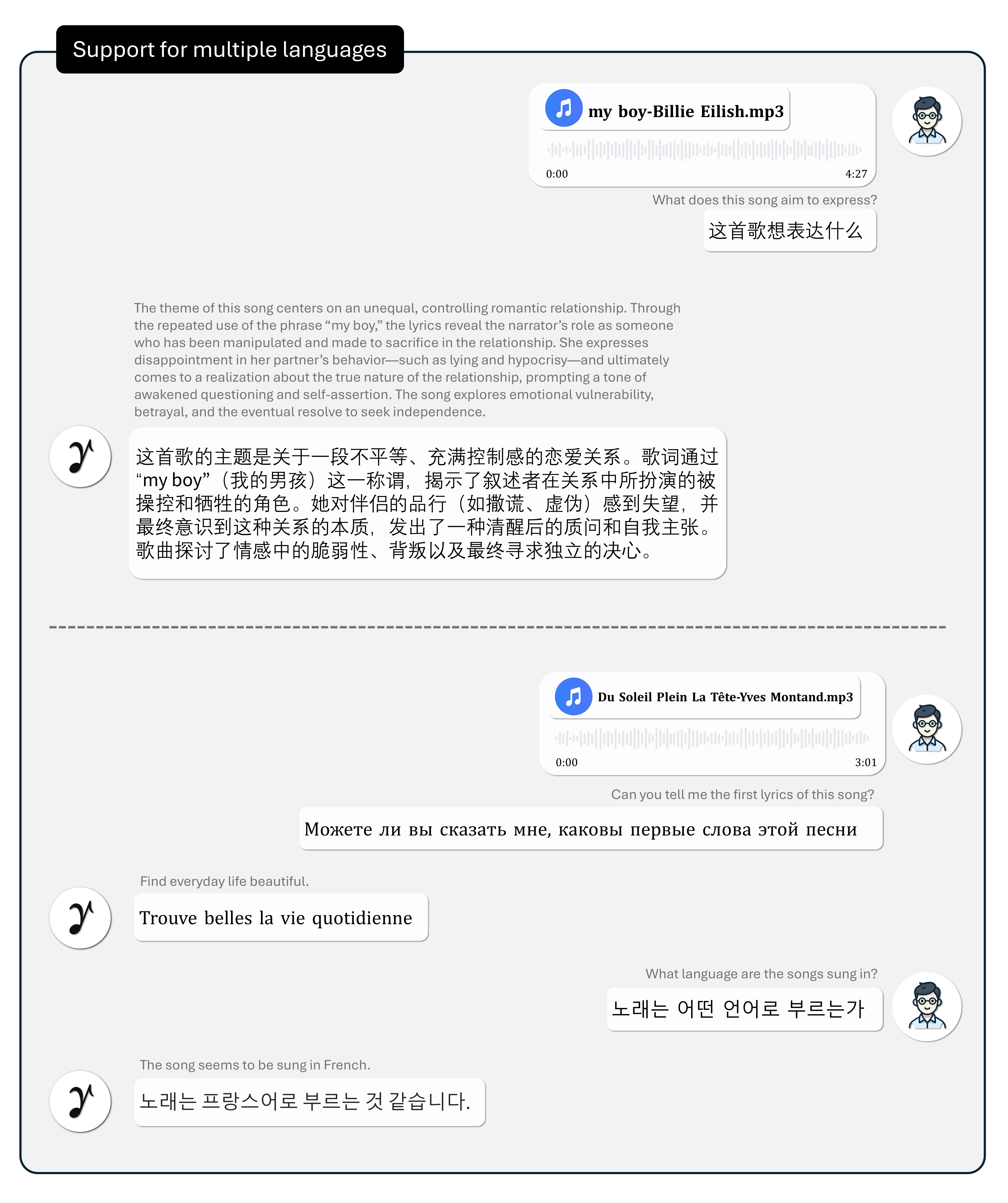}
  \caption{GaMMA is capable of understanding multiple languages and supports both multilingual input and output.}
  \label{multilingual}
\end{figure*}

\begin{figure*}[h]
  \centering
  \includegraphics[width=0.95\linewidth]{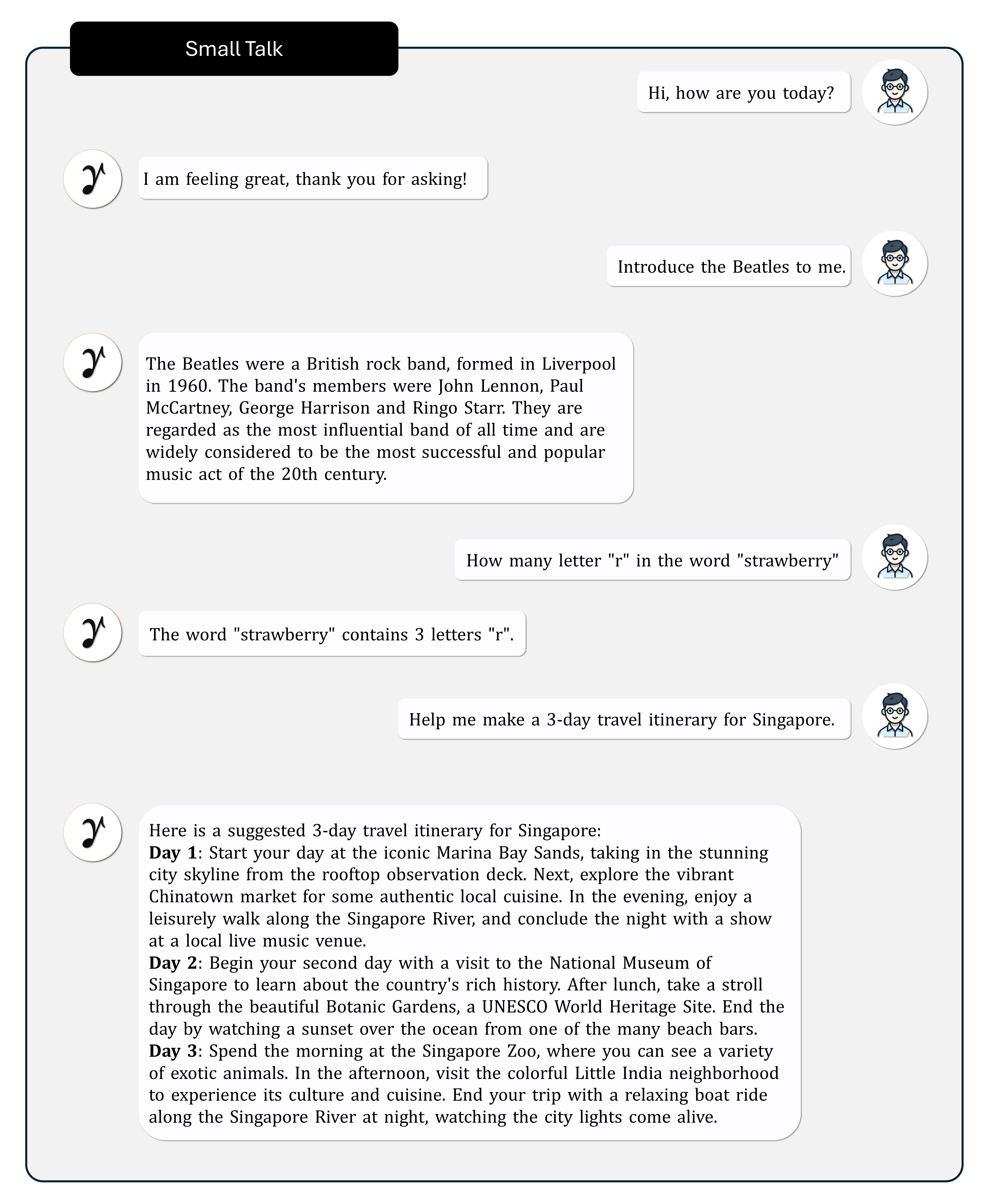}
  \caption{While designed as a music assistant, GaMMA demonstrates versatility beyond music-focused tasks.}
  \label{smalltalk}
\end{figure*}

We present the qualitative results of GaMMA in Fig.~\ref{sample1}–\ref{smalltalk}. To ensure that GaMMA truly understands music rather than merely memorizing training data, we tested it on several newly released songs not seen during training. As shown in the figures, GaMMA is able to generate highly detailed and accurate reports for each track, covering a wide range of musical aspects. Beyond analysis, it engages in coherent, content-grounded musical conversations, accurately interpreting emotional tone, theme, instrument, and lyrics. Furthermore, as shown in Fig.~\ref{multilingual} and Fig.~\ref{smalltalk}, GaMMA supports both multilingual input and output and casual dialogue, highlighting its strong potential as a versatile and general-purpose music assistant.

\begin{table*}[htbp]
\centering
\caption{QA samples from MusicBench}
\label{tab:musicbench_details}
\scriptsize
\begin{tabular}{p{2.5cm} p{10cm}}
\toprule

\small\textbf{Category} & \small\textbf{Question} \\
\midrule
\multicolumn{2}{l}{\small\textbf{\textit{MusicBench-Global}}} \\
\addlinespace
Primary Genre Label &What's the genre of this music? Options: (A) Epic (B) Devotional (C) Pop (D) Contemporary Chinese Music/Modern Chinese Music \\
\addlinespace
Secondary Genre Label &What's the genre of this music? Options: (A) Hardcore Punk (B) Nu Metal (C) Industrial Metal (D) City Pop \\
\addlinespace
Mood &What's the mood of this music?  Options: (A) Sorrow (B) Calm (C) Happy (D) Angry  \\
\addlinespace
Solo Instrument &What's the solo instrument of this music?  Options: (A) DiZi (B) Guzheng (C) Acoustic Piano (D) Violin
 \\
\addlinespace
Lead Instrument &What's the lead instrument of this music?  Options: (A) Music Box (B) DiZi (C) Violin (D) Guzheng
 \\
\addlinespace
Main Accompanying Instrument &What's the main accompanying instrument of this music?  Options: (A) Acoustic Piano (B) Guzheng (C) Xun (D) Sheng
 \\
\addlinespace
Vocal Attribute (Gender) &What's the vocal attribute (gender) of this music?  Options: (A) Child (B) Neutral (C) Male (D) Female \\
\addlinespace
Vocal Attribute (Timbre) &What's the vocal attribute (timbre) of this music?  Options: (A) Deep (B) Loud and sonorous (C) Warm (D) Sharp
 \\
\addlinespace
Vocal Attribute (Artist) &What's the vocal attribute (artist) of this music?  Options: (A) Rihanna (B) Mariah Carey (C) Lady Gaga (D) John Legend
 \\
\addlinespace
Language &What's the language of this music?  Options: (A) Cantonese (B) English (C) Chinese (D) French
 \\
\addlinespace
Contains Musical Instruments &What's the contains musical instruments of this music?  Options: (A) Piccolo (B) Tube (C) Wind Chimes (D) Oboe
 \\
\addlinespace
Key &What key is this song in?  Options: (A) C major (B) D minor (C) B minor (D) A major \\
\addlinespace
BPM &What is the BPM of this song?  Options: (A) 120BPM (B) 130BPM (C) 95BPM (D) 80BPM \\
\addlinespace
Melody &What is the melodic progression of this song?  Options: (A) Repetition (B) Variation (C) Truncation (D) Sequencing \\
\addlinespace
Rhythm &What time signature does this song have?  Options: (A) 6/8 (B) 7/8 (C) 4/4 (D) 3/8
 \\
\addlinespace
Content Association &What sort of scene or mood does this song bring about in your imagination?  Options: (A) The astronauts discovered a strange planet and began to explore it. (B) Several students were playing and joking around on the roadside. (C) A woman was sitting by the lake, staring at the water in a daze. (D) A person was drinking coffee and reading a book in a coffee shop.
 \\
\addlinespace
Scenario &What's the scenario of this music?  Options: (A) Rainy (B) Beach (C) Winter (D) Autumn \\
\addlinespace

\multicolumn{2}{l}{\small\textbf{\textit{MusicBench-Temporal}}} \\
\addlinespace

Lyrics &Which lyrics are heard from 0s to 6s?   Options: (A) But I still can't forget if I wanted too (B) But it's all in my head (C) I'm the first to say that I'm not perfect (D) I remember the day, Even wrote down the date, that I fell for you \\
\addlinespace
Chords &Identify the chord progression playing from 17s to 30s.   Options: (A) G:maj-D:maj-E:min7-C:maj7-D:maj (B) C:maj-D:maj-G:maj (C) D:7-C:7-G:7-G:7 (D) G:maj-D:maj-B:min-C:maj7-D:maj \\
\addlinespace
Structure &Which type of song structure does the 0s-14s of this song correspond to?   Options: (A) Bridge (B) Intro (C) Prechorus (D) Interlude \\
\addlinespace
Vocal &What is the gender of the voice at the 5 to 7 second mark in the designated audio section?  Options: (A) Female (B) Neutral (C) Child (D) Male \\
\addlinespace
Instrument &What is the solo instrument that appears between 0 to 10 second in the target audio section?   Options: (A) Acoustic Piano (B) Violin (C) Pipa (D) Guzheng \\

\bottomrule
\end{tabular}
\end{table*}

\begin{table}[htbp]
    \centering
    \renewcommand{\arraystretch}{1.5}
    \caption{Evaluation rubrics for expert study.}
    \scriptsize
    \begin{tabularx}{\textwidth}{@{}l X X@{}}
    \toprule
        
        \textbf{Category} & \textbf{Question} & \textbf{Rubrics} \\
        \midrule
        \multirow{2}{*}{\textbf{Vocal, language}} & \multirow{2}{=}{Describe the vocal information of this song, including whether it has vocals, the gender of the vocals, timbre, singing style, how many people are singing, In what language is the song performed, etc.} & 4: The output is complete and accurately identifies all major vocal attributes (gender / timbre / artist features) with no omissions or errors; or the result is error-free and is the optimal answer (e.g., the only reasonable option). \\
         & & 3: The output is error-free but does not fully cover all major vocal attributes (reasonable omissions exist); or the result is error-free but is not the most comprehensive answer (e.g., answers only part of the correct information). \\
         & & 2: The output contains some correct vocal attributes but includes incorrect information (e.g., wrong gender, timbre, or artist); or correct and incorrect attributes coexist. \\
         & & 1: The output completely fails to identify any correct vocal attributes (gender / timbre / artist features are all incorrect). \\
        \cmidrule(l){3-3}
         & & 4: The output is complete and accurately identifies all correct language features of the song (no omissions, no errors). \\
         & & 3: The output is error-free but does not fully list all language features (reasonable omissions exist). \\
         & & 2: The output contains some correct language features but includes incorrect languages (correct and incorrect coexist). \\
         & & 1: The output completely fails to identify any correct language features, or is irrelevant to the actual language (all information is incorrect). \\
        \midrule
        \textbf{Genre} & Introduce the genre of this song and analyze its primary genre and blended elements. & 4: The output precisely matches the most appropriate genre label for the song (i.e., ``optimal solution'') and fits the core characteristics of the genre (e.g., fine-grained classification). \\
         & & 3: One of the following occurs: (1) The output is correct but misses the optimal genre (chooses a suboptimal solution, e.g., ``indie pop''); (2) The output is error-free but only provides a coarse-grained genre (e.g., ``pop''). \\
         & & 2: The output contains some correct genres but includes incorrect labels (correct and incorrect genres coexist, or cross-category misjudgment). \\
         & & 1: The output completely deviates from the actual genre of the song and is unrelated to any correct genre (e.g., wrongly classified into an unrelated major genre category). \\

        \midrule
        \textbf{Mood} & Please describe the mood of this song. & 4: The output accurately identifies the most fitting mood tag (i.e., ``optimal solution'') without deviating from the core emotion. \\

         & & 3: The output aligns with the general emotional direction (falling within the correct category) but fails to identify the most precise mood tag (choosing a sub-optimal solution). \\

         & & 2: The output falls within the same emotional category (Positive/Negative/Neutral) as the correct answer, but the specific tag does not match the song's mood (tenuous or overly generic). \\

         & & 1: The output completely deviates from the actual emotion of the song and falls into a conflicting emotional category compared to the correct answer (e.g., Positive $\to$ Negative, Neutral $\to$ Extreme). \\

        \bottomrule
    \end{tabularx}
    \label{tab:rubric_1}
\end{table}

\begin{table}[htbp]
    \centering
    \renewcommand{\arraystretch}{1.5}
    \caption{Evaluation rubrics for expert study (continue).}
    \scriptsize
    \begin{tabularx}{\textwidth}{@{}l X X@{}}
        \toprule
        
        \textbf{Category} & \textbf{Question} & \textbf{Rubrics} \\

         \midrule
        \textbf{Theme/Scenario/Association} & What is the theme of this song? & 4: Completely correct. The output perfectly matches the ground truth without any errors. \\

         &In what scenarios is this song suitable for playback/listening? & 3: Acceptable. The output is error-free but does not represent the optimal answer. \\

         &What kind of plot or imagery does this song evoke? & 2: Partially correct. The output contains some correct information but includes significant errors.  \\

         & & 1: Completely incorrect. The output is entirely inconsistent with the song's actual characteristics and lacks any reasonable justification. \\

        \midrule
        \textbf{Instrument} & What instruments are used in this song? List them by lead instruments, harmonic instruments, bass instruments, and percussion instruments.& 4: The output completely and accurately identifies all instruments (without omissions or errors). \\

         & & 3: The output is error-free but may contain omissions. \\

         & & 2: The output identifies some correct instruments but includes incorrect ones. \\

         & & 1: The output completely fails to identify any correct instruments. \\

        \midrule
        \textbf{Highlight} &What is the biggest highlight of this song—the factor that made it go viral?
        & 4: Provides clear and unique details, such as specific instrument names or playing techniques (e.g., acoustic guitar arpeggios), specific vocal techniques or timbres (e.g., opera-style vocals, husky timbre), detailed descriptions of genre fusion (e.g., dialect rap mixed with hip-hop), specific examples or contrasts (e.g., explosiveness in the climax), and specific cultural or arrangement characteristics (e.g., pure piano accompaniment, J-Rock elements). \\

         & & 3: Provides more explicit information, such as specific emotions (e.g., longing, nostalgia), specific sound effects or processing (e.g., electronic effects, harmonies), general instrument categories or styles (e.g., traditional instruments, acoustic), specific functions or effects (e.g., suitable for dancing, warm atmosphere), and adjectives with some distinctiveness (e.g., full-bodied, soulful). \\

         & & 2: Uses common, non-specific adjectives (e.g., powerful, emotional, energetic, charming, captivating, moving) or describes the basic functions of elements (e.g., the voice carries the melody). \\

         & & 1: Merely mentions elements (e.g., vocals, rhythm) or uses only the most basic, generic positive descriptors (e.g., pleasant, nice, beautiful). \\

        \midrule
        \textbf{Melody} &Analyze the melodic development and melodic range of this song.& 4: The output accurately matches the most typical melodic progression characteristics of the song (i.e., ``optimal solution'') and aligns with the core melodic contour. \\

         & & 3: The output aligns with the actual melodic contour but fails to identify the most typical features (choosing a sub-optimal solution, non-core but reasonable); it must include a description of the melodic progression. \\

         & & 2: The output contains some correct features but includes incorrect descriptions (correct and incorrect features coexist, or deviation from the core melodic contour). \\

         & & 1: The output completely deviates from the actual melodic contour of the song and is unrelated to any correct features (e.g., the description is irrelevant or contrary to the melodic progression). \\

        \bottomrule
    \end{tabularx}
    \label{tab:rubrics_2}
\end{table}

\begin{table}[htbp]
    \centering
    \renewcommand{\arraystretch}{1.5}
    \caption{Evaluation rubrics for expert study (continue).}
    \scriptsize
    \begin{tabularx}{\textwidth}{@{}l X X@{}}
        \toprule
        
        \textbf{Category} & \textbf{Question} & \textbf{Rubrics} \\

        \midrule
        \textbf{Rhythm} &Analyze the rhythm, time signature, and groove of this song. & 4: The output completely and accurately identifies all distinct rhythm types present in the song, without omissions or errors. \\

         & & 3: The output is error-free, but one of the following occurs: (1) It does not fully list all distinct rhythms (reasonable omissions exist); (2) The description is coarse-grained (using general terms instead of specific rhythm types). \\

         & & 2: One of the following occurs: (1) Partially correct rhythms coexist with incorrect ones (e.g., inclusion of irrelevant types or excessive subdivision); (2) The output is related to the ground truth but the description is tenuous or inaccurate. \\

         & & 1: The output completely fails to identify any correct rhythms, or the description is unrelated to the song's actual rhythm (e.g., over-generalization, misclassification, or complete deviation). \\

        \midrule
        \textbf{Tempo/Key/Chord} &Please list the tempo, key, and chord progression in format of ``MM:SS.s chord1 - MM:SS.s chord2 - MM:SS.s chord3 - ...'' & 1. Annotation: For each song, record (1) the number of chord labels with errors in [root] or [majmin] ($E_{base}$), (2) the number of labels with errors in [root], [majmin], or [seventh] ($E_{full}$), and (3) the total number of chord labels ($N_{total}$). \newline
         2. Calculation: The accuracy metrics are calculated as $\text{Accuracy}_{base} = (N_{total} - E_{base}) / N_{total}$ and $\text{Accuracy}_{full} = (N_{total} - E_{full}) / N_{total}$. \newline
         3. Pass Criteria: The song passes if $\text{Accuracy}_{base} \ge 90\%$ and $\text{Accuracy}_{full} \ge 80\%$. \\

        \midrule
        \textbf{First verse–chorus progression} &Analyze the energy trend of the first verse-chorus group. Describe the energy changes in the first verse-chorus group, such as starting subdued then building up, starting strong then tapering off, or maintaining consistent energy.  & 4: Completely correct. The output is semantically identical to the ground truth, accurately describing the direction and intensity of the energy trend. Variations in phrasing are permissible, provided the core information is equivalent to the ground truth. \newline

         3: Acceptable. The output is error-free but does not represent the optimal answer. \newline

         2: Partially correct. The output contains some correct information but includes significant errors. \newline

         1: Completely incorrect. The output is entirely inconsistent with the energy dynamic relationship between the verse and chorus, lacking any reasonable justification. \\

        \midrule
        \textbf{Structure} &Please provide a detailed analysis of this song's structure, including the time range of each section and the lyrics (if any) under each musical structure. For each section, List the instruments, the song's time signature, the rhythmic feel, melodic information.
        & 4: The [Section Labels] in the output are completely correct and identical to the ground truth, AND the [Timestamp Boundaries] error for each section is $\le \pm 1$s. \newline

         3: The [Section Labels] are error-free but may not represent the optimal annotation method, AND the [Timestamp Boundaries] error for each section is $\le \pm 2$s. \newline

         2: Some [Section Labels] are incorrect, OR the [Timestamp Boundaries] error for some sections is $> \pm 2$s. \newline

         1: All [Section Labels] are incorrect, OR the [Timestamp Boundaries] error for all sections exceeds $\pm 2$s. \\

        \bottomrule
    \end{tabularx}
    \label{tab:rubrics_3}
\end{table}

